# Don't "research fast and break things":
# On the ethics of Computational Social Science[*]

David Leslie
The Alan Turing Institute
**dleslie@turing.ac.uk**

## Abstract

As a quintessential *social impact science*, Computational Social Science (CSS) holds great promise to advance social justice, human flourishing, and biospheric sustainability. However, CSS is also an all-too-human science—conceived in particular social, cultural, and historical contexts and pursued amidst intractable power imbalances, structural inequities, and potential conflicts of interest. Its proponents must thus remain continuously self-critical about the role that values, interests, and power dynamics play in shaping mission-driven research. Likewise, they must take heed of the complicated social and historical conditions surrounding the generation and construction of data as well as the way that the activities and theories of CSS researchers can function to restructure and shape the phenomena that they purport only to measure and analyse. This article is concerned with setting up practical guardrails within the research activities and environments of CSS in response to these dilemmas. It aims to provide CSS scholars, as well as policymakers and other stakeholders who apply CSS methods, with the critical and constructive means needed to ensure that their practices are ethical, trustworthy, and responsible. It begins by providing a taxonomy of the ethical challenges faced by researchers in the field of CSS. These are challenges related to (1) the treatment of research subjects, (2) the impacts of CSS research on affected individuals and communities, (3) the quality of CSS research and to its epistemological status, (4) research integrity, and (5) research equity. Taking these challenges as a motivation for cultural transformation, it then argues for the end-to-end incorporation of habits of responsible research and innovation (RRI) into CSS practices, focusing on the role that contextual considerations, anticipatory reflection, impact assessment, public engagement, and justifiable and well-documented action should play across the research lifecycle. In proposing the inclusion of habits of RRI in CSS practices, the chapter lays out several practical steps needed for ethical, trustworthy, and responsible CSS research activities. These include stakeholder engagement processes, research impact assessments, data lifecycle documentation, bias self-assessments, and transparent research protocols.

[*] This paper is an unabridged pre-print of a chapter written for the European Commission's Joint Research Centre, Scientific Development Centre for Advanced Studies, to be published in *Handbook of Computational Social Science for Policy* (2022) by *Springer*. In addition to the JRC's support, the author would like to acknowledge the support of a grant from ESRC (ES/T007354/1), Wave 1 of The UKRI Strategic Priorities Fund under the EPSRC Grant EP/W006022/1, Towards Turing 2.0 under the EPSRC Grant EP/W037211/1, and the public funds that make the Turing's Public Policy Programme possible. The author would additionally like to thank Serena Signorelli, Claudia Fischer, and Morgan Briggs for their invaluable editorial assistance.

# Introduction: Combatting the lures of scientism in CSS

This paper is concerned with setting up practical guardrails within the research activities and environments of Computational Social Science (CSS). It aims to provide CSS scholars with the critical and constructive means needed to ensure that their research is ethical, trustworthy, and responsible. This may seem like an obvious first priority for any scholarly community of practice that involves human subjects in research and that impacts individuals and communities with the results, capabilities, and tools it generates. However, over the relatively short history of CSS, this has evidently not been the case. Across the four volumes of Nigel Gilbert's magisterial *Computational Social Science* (2010), none of the 66 contributing chapters are dedicated to ethics. Likewise, no explicit mention or discussion of research ethics appears in Conte et al.'s programmatic and widely cited "Manifesto of computational social science," published in 2012. There are only two passing mentions of ethics in the ten chapters of Cioffi-Revilla's substantial *Introduction to Computational Social Science* (2014), and the word "ethics" also appears only twice (and only in the final chapter) of Chen & Yu's edited volume, *Big Data for the Computational Social Sciences and Humanities* (2018).

Though some significant attempts to articulate the ethical stakes of CSS have been made by scholars and professional associations over the past two decades,[1] the scarcity of ethics in the mainstream labours of CSS, and across its history, signals a general lack of awareness that is illustrative of several problematic dimensions of current CCS research practices that will motivate the arguments presented in this chapter. It is illustrative insofar as the absence of an active recognition of the ethical issues surrounding the social practice and wider human impacts of CSS may well shed light on a troublesome strain of scientism in its culture that requires urgent correction and transformation. As Sorell (2013) has argued, scientism is typified by the privileging of natural or exact scientific language, knowledge, and methods over those of other branches of learning and culture, especially those of the "human sciences" like philosophy, ethics, history, anthropology, and sociology. Such a privileging of exact scientific "ideas, methods, practices, and attitudes" can be especially damaging where these are extended "to matters of human social and political concern" (Olson, 2008, p. 1)—matters that require an understanding of subtle historical, ethical, and sociocultural contexts, contending human values, norms, and purposes, and subjective meaning-complexes of action and interaction (Apel, 1984; Habermas, 1967/1988; Taylor, 1964/2021; Von Wright, 1971/2004; Weber, 1922/1978; Wittgenstein, 1953/2009) .

In the case of CSS, the hazards of scientistic attitudes are more pronounced and consequential than in other more disciplinarily anchored fields, because its methodologies explicitly straddle the line between the exact, quantitative character of applied statistics, mathematical modelling, and complexity science and the more interpretive and qualitative character of the traditional social sciences.  Indeed, since its inception, one of the great promises of CSS has been the possibility of leveraging the former, computational side, to gain insights and identify patterns in big data that would have otherwise been unavailable to the latter, interpretive side—an advancement made possible in virtue of the potential of CSS to overcome what

---

[1] See, for instance, the series of AoIR guidelines on internet research ethics published in 2002, 2012, and 2019 as well as the BSA guidance. For scholarly interventions, see Collmann & Matei, 2016; Dobrick et al., 2018; Ess & Jones, 2004; Eynon et al., 2016; franzke et al., 2020; Giglietto et al., 2012; Hollingshead et al., 2021; Lomborg, 2013; Markham & Buchanan, 2012; Moreno et al., 2013; Salganik, 2019; Weinhardt, 2020.

Burrell (2016) has referred to as the "mismatch" between the mathematical complexity of high-dimensional computation and the limitations of "human-scale reasoning and styles of semantic interpretation". That is, as the aspirational story goes, through the application of a variety of computational and algorithmic techniques to the vast amounts of data generated from today's complex, digitised, and datafied society, CSS generates empirically grounded inferences, explanations, theories, and predictions about human behaviours, networks, and social systems which not only effectively manage the volume and high dimensionality of big social data but, in fact, draw epistemic advantage from such an unprecedented breadth, quantity, depth, and scale of data.

Mainly owing to these perceived cognitive advantages, computation-driven methods have, to a certain degree, functioned to ingrain the hydra-headed perils of scientism in CSS practices. The portfolio of such perils is worth exploring, for it illuminates some of the ethical deficits that are in need of remediation. To begin with, the scientistic prioritisation of computational approaches can lead to reductivism. Where CSS approaches appeal exclusively to the theoretical paradigm of complex adaptive systems and to the affiliated battery of behaviourist, game theoretic, cybernetic, and simulation-based modes of explanation, they can elide the considerable interpretive and semantic attentiveness needed to understand complex, pluralistic, ever-interacting, and ever-changing social norms and practices, human intentions, and cultural forms of life that condition the production, construction, and interpretation of data. None of these contextually and socio-culturally situated factors is, in a strict sense, reducible to computational formalisations, network analyses, or simulation-based heuristics (Burrell & Fourcade, 2021; Stark, 2018). In a similar vein, where CSS approaches displace the scientific method, relying instead on voluminous data and computational dexterity to generate high-performance forecasting (Anderson, 2008; Hindman, 2015; Lin 2015), they can reductively disregard this interpretive aspect entirely by replacing the epistemic aspirations of explanation and understanding with the deployment of algorithmic techniques that optimise for predictive accuracy alone (Hofman et al., 2021; Shah et al., 2015).

In each of these cases, scientistic reductivism in CSS can lead its proponents into prejudicial ditches. First, it can drive CSS researchers into speculative and largely unsupported metanarratives of progress wherein teleological accounts of supposed mathematical and computational paradigm shifts in the previously "humanistic disciplines" of the traditional social sciences like linguistics, economics, and sociology "propel" them "into the modern science[s] that [they are] today" (Cioffi-Revilla, 2014, p. 3).[2] Amidst the intellectual bounties of this sensed progress, proponents of such "modernised" computational approaches to the social sciences can presumptively leave behind the outmoded historiographical and hermeneutic sensibilities of bygone days. This metaphysical view of scientistic advancement, however, selectively avoids the other, more qualitatively driven and contextually responsive side of the development of social theory over the past several generations. Namely, it overlooks the evolution of deflationary and interpretive approaches to the study of society that have built on the linguistic, pragmatic, and post-positivist turns in the twentieth century human sciences and that now underwrite critical and constructive approaches to understanding complex social practices, power dynamics, and human agents in culturally sensitive, dialogically anchored, contextually responsive, and pluralistically informed ways (Bernstein, 1971; Groff, 2004; Sellars, 1956/1997; Taylor, 1973; Winch, 1958/2015; Zammito, 2004). Rather than having waned, these more "humanistic" approaches have gained strength in recent years, heavily influencing geological framings of the Anthropocene epoch (Bonneuil & Fressoz, 2016; Chakrabarty, 2018; Zinn, 2016), driving the sociotechnical and ELSA/ELSI self-understandings of the study of emerging technologies (Fisher, 2005; Hullmann, 2008; Rip, 2000; Swierstra & Rip, 2007), and growing within the Academy under the auspices of interdisciplinary and sub-disciplinary fields like Science and Technology Studies (Jasanoff, 2012, 2016; Sengers et al., 2005; Star, 1999), Critical Sociology (Bauman, 1976/2010; Burawoy, 1998; Bourdieu & Wacquant, 1992) , Responsible Research and Innovation (Hellström, 2003; Owen et al., 2012; Von Schomberg, 2011, 2013, 2019), Environmental Humanities (Emmett & Nye, 2017; Oppermann & Iovino, 2016), Critical Data Studies (Dalton et al., 2013; Iliadis & Russo, 2016; Kitchin & Lauriault, 2014), Postcolonial and Decolonial Theory (de Sousa Santos, 2014; Grosfoguel et al., 2007; Mohamed et al., 2020), and Cultural Studies (Barker, 2003; Johnson et al., 2004), among others.

---

[2] Hofman et al. (2021), similarly though more modestly, observe that, "in the past 15 years, social science has experienced the beginnings of a 'computational revolution' that is still unfolding" (p. 181).

Largely steering clear of this universe of research and insight has, in a significant sense, led some scientistic corners of CSS research into a cramped and selective form of interdisciplinarity where its inclusion criteria tends to be based on quantitative affinities, computational predispositions, and conceptual proximity to the "information processing" or "computational paradigm of society" (Cioffi-Revilla, 2014, p. 2) (Diebel-Fischer, 2018). For instance, Conte et al. (2012) declare that CSS "is a truly interdisciplinary approach, where social and behavioural scientists, cognitive scientists, agent theorists, computer scientists, mathematicians and physicists cooperate side-by-side to come up with innovative and theory-grounded models of the target phenomena" (p. 327). The hazard here is that an implicit orientation to selective interdisciplinarity can lead to a kind of self-serving academic virtue signalling about the embrace of interdisciplinary research rather than an ongoing pursuit of authentic cross-disciplinary initiatives and programmes that integrate critical perspectives and alternative approaches.

Another related prejudicial ditch that imperils the ethics of CSS research has to do with objectification. Scientistic reductivism in CSS can lead researchers to assume positivistic attitudes that frame the objects of their study through the quantifying and datafying lenses of models, formalisms, behaviours, networks, simulations, and systems thereby setting aside or trivialising ethical considerations in an effort to get to the real science without further ado. When the *objects* of the study of CSS are treated solely as elements of automated information analysis rather than as *human subjects*—each of whom possesses a unique dignity and is thus, first and foremost, worthy of moral regard and interpretive care—scientistic subspecies of CSS are liable to run roughshod over fundamental rights and freedoms like privacy, autonomy, meaningful consent, and non-discrimination with a blindered view to furthering computational insight and data quantification (Fuchs, 2018; Hollingshead et al., 2021).

This potentially objectifying stance can also have direct recoil effects on the self-understanding of CSS researchers who implicitly see themselves more as neutral and disinterested scientists operating within the pure, self-contained confines of the laboratory or lecture hall than as contextually situated scholars whose categories, subject matters, framings, and methods have been forged in the crucible of history, society, and culture. To be sure, when CSS researchers assume a scientistic "view from nowhere" and regard the objects of their study solely through quantitative and computational lenses, they run the risk of seeing themselves as operationally independent or even immune from the conditioning dynamics of the social environments they study and in which their own research activities are embedded (Feenberg, 2002, 2012). This can create conditions of deficient reflexivity—i.e. defective self-awareness of the limitations of one's own standpoint—and ethical precarity (Leslie et al., 2022a). As John Dewey long ago put it, "the notion of the complete separation of science from the social environment is a fallacy which encourages irresponsibility, on the part of scientists, regarding the social consequences of their work" (Dewey, 1938, p. 489).

In the case of CSS, the price of this misperceived independence of researchers from the formative dynamics of their sociohistorical environment has been extremely high. CSS practices have developed and matured in an age of unprecedented sociotechnical sea change—an age of unbounded digitisation, datafication, and mediatization. The cascading societal effects of these revolutionary transformations have, in fact, directly shaped and implicated CSS in its research trajectories, motivations, objects, methods, and practices. The rise of the veritably limitless digitisation and datafication of social life has brought with it a corresponding impetus—among an expanding circle of digital platforms, private corporations, and governmental bodies—to engage in behavioural capture and manipulation at scale. In this wider societal context, the aggressive extraction and harvesting of data from the digital streams and traces generated by human activities, more often than not, occurs without the meaningful consent or active awareness of the people whose digital and digitalised lives[3] are the targets of increasing surveillance, consumer curation, computational herding, and behavioural steering. Such extractive and manipulative uses of computational technologies also often occur neither with adequate reflection on the potential transformative effects that they could have on the identity formation, agency, and autonomy of targeted data subjects nor with appropriate and community-involving assessment of the adverse impacts they could have on civic and social freedoms, human rights, the integrity of interpersonal relationships, and communal and biospheric wellbeing. The real threat here, for CSS, is that the prevailing "move fast and break things" attitude possessed by the drivers of the "big data revolution", and by the beneficiaries of its financial and administrative windfalls, will simply be transposed into the key of the data-driven research practices they influence, making a "research fast and break things" posture in CSS a predominant disposition. This threat to the integrity of CSS research activity, in fact, derives from the potentially inappropriate dependency relationships which can emerge from power imbalances that exist between the CSS community of practice and those platforms, corporations, and public bodies who control access to the data resources, compute infrastructures, project funding opportunities, and career advancement prospects upon which CSS researchers rely for their professional viability and endurance. Here, the misperceived independence of researchers from their social environments can mask toxic and agenda-setting dependencies.

---

[3] The use of the the terms 'digital' and 'digitalised' follows Lazer & Radford, (2017).

Taken together, these downstream hazards of scientistic orientations in CSS signal potential deficits in the social responsibility, trustworthiness, and ethical permissibility of its practices. To confront such hazards, this chapter will first provide a taxonomy of ethical challenges faced by CSS researchers. These are (1) challenges related to the treatment of research subjects, (2) challenges related to the impacts of CSS research on affected individuals and communities, (3) challenges related to the quality of CSS research and to its epistemological status, (4) challenges related to research integrity, and (5) challenges related to research equity. Taking these challenges as a motivation for cultural transformation, it will then argue for the end-to-end incorporation of habits of responsible research and innovation (RRI) into CSS practices, focusing, in particular, on the role that contextual considerations, anticipatory reflection, public engagement, and justifiable action should play across the research lifecycle. The primary goal of this focus on RRI is to centre the understanding of CSS as "science with and for society" and to foster, in turn, critical self-reflection about the consequential role that human values, norms, and purposes play in its discovery and design processes and in considerations of the real-world effects of the insights and tools that these processes yield. Finally, the chapter will conclude by proposing several practical steps needed for ethical, trustworthy, and responsible CSS research practices. These include stakeholder engagement processes, research impact assessments, data lifecycle documentation, bias self-assessments, and transparent research reporting protocols.

## Ethical challenges faced by CSS

A preliminary step needed to motivate the centring of responsible research and innovation practices in CSS is the identification of the range of ethical challenges faced by its researchers. These challenges can be broken down into five categories:

1. **Challenges related to the treatment of research subjects.** These challenges have to do with the interrelated aspects of confidentiality, data privacy and protection, anonymity, and informed consent.

2. **Challenges related to the impacts of CSS research on affected individuals and communities.** These challenges cover areas such as the potential adverse impacts of CSS research activities on the respect for human dignity and on other fundamental rights and freedoms. This includes the potential transformative effects that CSS research practices, outcomes, and tools could have on individual agency and identity formation, physical, psychological, and moral integrity, interpersonal connection and social solidarity, and personal, communal, and biospheric wellbeing.

3. **Challenges related to the quality of CSS research and to its epistemological status.** Challenges related to the quality of CSS research include erroneous data linkage, dubious "ideal-user assumptions", the infusion of algorithmic influence in observational datasets of digital traces, the 'illusion of the veracity of volume', and blind spots vis-à-vis non-human data generation that undermine data quality and integrity. Challenges related to the epistemological status of CSS include the inability of computation-driven techniques to fully capture non-random missingness in datasets and sociocultural conditions of data generation and hence a broader tendency to potentially misrepresent the real social world in the models, simulations, analyses, and predictions it generates.

4. **Challenges related to research integrity.** These challenges are rooted in the asymmetrical dynamics of resourcing and influence that can emerge from power imbalances between the CSS research community and the corporations and public agencies upon whom CSS scholars rely for access to the data resources, compute infrastructures, project funding opportunities, and career advancement prospects they need for their professional subsistence and advancement.

5. **Challenges related to research equity.** These challenges include the potential reinforcement of digital divides and data inequities through biased sampling techniques that render digitally marginalised groups invisible as well as potential aggregation biases in research results that mask meaningful differences between studied subgroups and therefore hide the existence of real-world inequities. Research equity challenges may also derive from long-standing dynamics of regional and global inequality that may undermine reciprocal sharing between research collaborators from more and less resourced geographical areas, universities, or communities of practice.

Let us expand on each of these challenges in turn.

## Challenges related to the treatment of research subjects.

When identifying and exploring challenges related to the treatment of research subjects in CSS, it is helpful to make a distinction between participation-based and observation-based research, namely, between CSS research that is gathering data directly from research subjects through their deliberate involvement in digital media (for example, research that uses online methods to gather data by way of human involvement in surveys, experiments, or participatory activities) and CSS research that is investigating human action and social interaction in observed digital environments, like social media or search platforms, through the recording, measurement, and analysis of digital life, digital traces, and digitalised life (Eynon et al., 2016). Though participation-based and observation-based research raise some overlapping issues related to privacy and data protection, there are notable differences that yield unique challenges.

Several general concerns about privacy preservation, data protection, and the responsible handling and storage of data are common to participation-based and observation-based CSS research. This is because empirical CSS research often explores topics that require the collection, analysis, and management of personal data, i.e., data that can uniquely identify individual human beings. The European Union's General Data Protection Regulation (GDPR)[4]—which is widely regarded as a gold standard for ensuring responsible data collection, management, and use—safeguards particular categories of personal data that are deemed especially sensitive. These include personal data revealing racial or ethnic origin, political opinions, religious or philosophical beliefs, trade union membership, genetic and biometric data, and data concerning health, sexual life, and person's sexual orientation (GDPR, art. 9; ICO, 2021). All forms of CSS research that use sensitive data or data which could allow for the identification of natural persons, whether such persons have consented to the processing or not, fall under obligations to handle and manage this data ethically and lawfully (Weinhardt, 2020) from beginning to end of their research process (Nosek et al., 2002), so to prevent harms that could occur because of breaches of confidentiality and anonymity triggered by inappropriate data collection, handling, management, or use (Eynon et al., 2016; Fox et al., 2003). Although CSS research frequently spans different jurisdictions, which may have diverging privacy and data protection laws, responsible research practices that aim to optimally protect the rights and interests of research subjects in light of risks posed to confidentiality, privacy, and anonymity should recur to the highest standards of privacy preservation, data protection, and the responsible handling and storage of data. They should also establish and institute proportionate protocols for attaining informed and meaningful consent that are appropriate to the specific contexts of the data extraction and use and that cohere with the reasonable expectations of the targeted research subjects.

Notwithstanding this common footing for ethics considerations related to data protection and the privacy of research subjects, participation-based and observation-based approaches to CSS research each raise distinctive issues. For researchers who focus on online observation or who use data captured from digital traces or data extracted from connected mobile devices, the Internet of Things, public sensors and recording devices, or networked cyber-physical systems, coming to an appropriate understanding of the reasonable expectations of research subjects regarding their privacy and anonymity is a central challenge. When observed research subjects move through their synchronous digital and connected environments striving to maintain communication flows and coherent social interactions, they must navigate moment-to-moment choices about the disclosure of personal information (Joinson et al., 2007). In physical public spaces and in online settings, the perception of anonymity (i.e. of the ability to speak and act freely without feeling like one is continuously being identified or under constant watch) is an important precondition of frictionless information exchange and, correspondingly, of the exercise of freedoms of movement, expression, speech, assembly, and association (Jiang et al., 2013; Paganoni, 2019; Selinger & Hartzog, 2020). On the internet, moreover, an increased sense of anonymity may lead data subjects to more freely disclose personal information, opinions, and beliefs that they may not have shared in offline milieus (Meho, 2006). In all these instances of perceived anonymity, research subjects may act under reasonable expectations of gainful obscurity and 'privacy in public' (Nissenbaum, 1998; Reidenberg, 2014). These expectations are responsive to and bounded by the changing contexts of communication, namely, by contextual factors like who one is interacting with, how one is exchanging information, what type of information is being exchanged, how sensitive it is perceived to be, and where and when such exchanges are occurring (Quan-Haase & Ho, 2020). This means, not only, that the protection of privacy must, first and foremost, consider contextual determinants (Collmann & Matei, 2016; Nissenbaum, 2011; Steinmann et al. 2015). It also implies that privacy protection considerations must acknowledge that the privacy preferences of research subjects can change from circumstance to circumstance and are therefore not one-off or one-dimensional decisions that can be made at the entry point to the usage of digital or social media applications through Terms of Service or end-user license agreements—which often go unread—or the initial determination of privacy settings (Henderson et al., 2013). For this reason, the conduct of observation-based research in CSS that pertains to digital and digitalised life should be informed by contextual considerations about the populations and social groups from whom the data are drawn, the character and potential sensitivities of their data, the nature of the research question (as it may be perceived by observed research subjects), research subjects' reasonable expectations of privacy in public, and the data collection practices and protocols of the organisation or company which has extracted the data (Hollingshead et al., 2021). Notably, thorough assessment of these issues by members of a research team may far exceed formal institutional processes for gaining ethics approval, and it is the responsibility of CSS researchers to evaluate the appropriate scale and depth of privacy considerations regardless of minimal legal and institutional requirements (Eynon et al., 2016; Henderson et. al., 2013).

Apart from these contextual considerations, the protection of the privacy and anonymity of CSS research subjects also requires that risks of re-identification through triangulation and data linkage are anticipated and addressed. While processes of anonymisation and removal of personally identifiable information from datasets scraped or extracted from digital platforms and digitalised behaviour may seem straightforward when those data are treated in isolation, multiple sources of linkable data points and multiple sites of downstream data collection pose tangible risks of re-dentification via the combination and linkage of datasets (de Montjoye et al., 2015; Eynon et al., 2016; Oboler et al., 2012). As Narayanan & Shmatikov (2009) and de Montjoye et al. (2015) both demonstrate, the inferential triangulation of social data collected from just a few sources can lead to re-identification even under conditions where datasets have been anonymised in the conventional, single dataset sense. Moreover, when risks of triangulation and re-identification are considered longitudinally, downstream risks of de-anonymisation also arise. In this case, the endurance of the public accessibility of social data on the internet over time means that information that could lead to re-identification is ready-to-hand indefinitely. By the same token, the production and extraction of new data that post-dates the creation and use of anonymised datasets also present downstream opportunities for data linkage and inference creep that can lead to re-identification through unanticipated triangulation (Weinhardt, 2020).

Although many of these privacy and data protection risks also affect participation-based research (especially in cases where observational research is combined or integrated with it), experimental and human-involving CSS projects face additional challenges. Signally, participation-based CSS research must confront several issues surrounding the ascertainment of informed and meaningful consent. The importance of consent has been a familiar part of the "human subjects" paradigm of research ethics from its earliest expressions in the WMA Declaration of Helsinki[5] and the Belmont Report[6]. However, the exponentially greater scale and societal penetration of CSS in comparison to more conventional forms of face-to-face, survey-driven, or laboratory-based social scientific research present a new order of hazards and difficulties. First, since CSS researchers, or their collaborators, often control essential digital infrastructure like social media platforms, they have the capability to efficiently target and experiment on previously unimaginable numbers of human subjects, with potential N's approaching magnitudes of hundreds of thousands or even millions of people. Moreover, in the mould of such platforms, these researchers have an unprecedented capacity to manipulate or surreptitiously intervene in the unsuspecting activities and behaviours of such large, targeted groups.

The controversy around the 2014 Facebook emotional contagion experiment demonstrates some of the potential risks generated by this new scale of research capacity (Grimmelmann, 2015; Lorenz, 2014; Puschmann & Bozdag, 2014). In the study, researchers from Facebook, Cornell, and the University of California involved almost 700,000 unknowing Facebook users in what has since been called a "secret mood manipulation experiment" (Meyer, 2014). Users were split into two experimental groups and exposed to negative or positive emotional content to test whether News Feed posts could spread the relevant positive or negative emotion. Critics of the approach soon protested that the failure to obtain consent—or even to inform research subjects about the experiment—violated basic research ethics. Some also highlighted the dehumanising valence of these research tactics: "To Facebook, we are all lab rats," wrote Vindu Goel in the *New York Times* (Goel, 2014). Hyperbole aside, this latter comment makes explicit the internal logic of many of the moral objections to the experiment that were voiced at the time. The Facebook researchers had blurred the relationship between the laboratory and the lifeworld. They had, in effect, unilaterally converted the social world of people connecting and interacting online into a world of experimental objects that subsisted merely as standing reserve for computational intervention and study—a transformation of the interpersonally animated life of the community into the ethically impoverished terrain of an "information laboratory" (Cohen, 2019a). Behind such a degrading conversion was the assertion of the primacy of objectifying and scientistic attitudes over considerations of the equal moral status and due ethical regard of research subjects. The experiment had, on the critical view, *reduced Facebook users to the non-human standing of laboratory rodents*, thereby disregarding their dignity and autonomy and consequently failing to properly consult them so to attain their informed consent to participate.

Even when the consent of research participants is sought by CSS researchers, a few challenges remain. These revolve around the question of how to ensure that participants are fully informed so that they can freely, meaningfully, and knowledgeably consent to their involvement in the research (franzke et al., 2019). Though diligent documentation protocols for gaining consent are an essential element of ascertaining informed and meaningful consent in any research environment, in the digital or online milieus of CSS, the provision of this kind of text-based information is often inadequate. When consent documentation is provided in online environments through one-way or vertical information flows that do not involve real, horizontal dialogue between researchers and potential research subjects, opportunities to clarify possible misunderstandings of the terms of consent can be lost (Varnhagen et al., 2005). What is more, it becomes difficult under these conditions of incomplete or impeded communication to confirm that research subject actually comprehend what they are agreeing to do as research participants (Eynon et al., 2016). Relatedly, barriers to information exchange in the online environment can prevent researchers from being able to verify the capacity of research subjects to consent freely and knowledgeably (Eynon et al., 2016; Kraut et al., 2004). That is, it is more difficult to detect potential limitations of or impairments in the competence of participants (e.g. from potentially vulnerable subgroups) in giving consent where researchers are at a significant digital remove from research subjects. In all these instances, various non-dialogical techniques for confirming informed consent are available—such as comprehension tests, smart forms that employ branching logic to ensure essential text is completely read, identity verification, etc. Such techniques, however, present varying degrees of uncertainty and drop-out risk (Kraut et al., 2004; Varnhagen et al., 2005), and they do not adequately substitute for interactive mechanisms that could connect researchers directly with participants and their potential questions and concerns.

## Challenges related to the impacts of CSS research on affected individuals and communities

While drawing on the formal techniques and methods of mathematics, statistics, and the exact sciences, CSS is a research practice that is policy-oriented, problem-driven, and societally consequential. As an applied science that directly engages with issues of immense social concern like socioeconomic inequality, the spread of infectious disease, and the growth of disinformation and online harm, it impacts individuals and communities with the results, capabilities, and tools it generates. Moreover, CSS is an "instrument-enabled science" (Cioffi-Revilla, 2013, p. 4) that employs computational techniques, which can be applied to large-scale datasets excavated from veritably all societal sectors and spheres of human activity and experience. This makes its researchers the engineers and custodians of a *general purpose research technology* whose potential scope in addressing societal challenges is seemingly unbounded. With this in view, Lazer et al. (2020) call for the commitment of "resources, from public and private sources, that are extraordinary by current standards of social science funding" to underwrite the rapid expansion of CSS research infrastructure, so that its proponents can enlarge their quest to "solve real-world problems" (p.1062). The authors continue,

> To justify such an outsized investment, computational social scientists must make the case that the result will be more than the publication of journal articles of interest primarily to other researchers. They must articulate how the combination of academic, industrial, and governmental collaboration and dedicated scientific infrastructure will solve important problems for society—saving lives; improving national security; enhancing economic prosperity; nurturing inclusion, diversity, equity, and access; bolstering democracy; etc. (Ibid).

Beyond the dedication of substantial resources, such an expansion, Lazer et al. (2020) argue, also requires the formulation of "policies that would encourage or mandate the ethical use of private data that preserves public values like privacy, autonomy, security, human dignity, justice, and balance of power to achieve important public goals—whether to predict the spread of disease, shine a light on societal issues of equity and access, or the collapse of the economy" (p. 1061). CSS, along these lines, is not simply an applied social science. It is a social impact science *par excellence.*

The mission-driven and impact-oriented perspective conveyed here is, however, a double-edged sword. On the one hand, the drive to improve the human lot and to solve societal problems through the fruits of scientific discovery has constructively guided the impetus of modern scientific research and innovation at least since the 17th century dawning of the Baconian and Newtonian revolutions. This sense of the humane vocation of science was already expressed by Francis Bacon in 1605, when he famously remarked that "science discovery should be driven not just by the quest for intellectual enlightenment, but also for the relief of man's estate". In this sense, the practical and problem-solving aspirations for CSS expressed by Lazer et al. (2020) are continuous with a deeper tradition of societally oriented science.

On the other hand, the view that CSS is a mission-driven and impact-oriented science raises a couple of thorny ethical issues that are not necessarily solvable by the application of its own methodological and epistemic resources. First, the assumption of a mission-driven starting point surfaces a difficult set of questions about the relationship of CSS research to the values, interests and power dynamics that influence the trajectories of its practice: *Whose* missions are driving CSS and w*hose* values and interests are informing the policies that are guiding these missions? To what extent are these values and interests shared by those who are likely to be impacted by the research? To what extent do these values and interests, and the policies they shape, sufficiently reflect the plurality of values and interests that are possessed by members of communities who will potentially be affected by the research (especially those from historically marginalised, discriminated-against, and vulnerable social groups)? Are these missions determined through democratic and community-involving processes or do other parties (e.g. funders, research collaborators, resource providers, principal investigators, etc.) wield asymmetrical agenda-setting power in setting the direction of travel for the research and its outputs? Who are the beneficiaries of these mission-driven research projects and who are at risk of any adverse impacts that they could have? Are these potential risks and benefits equitably distributed or are some stakeholders disparately exposed to harm while others in positions of disproportionate advantage?

Taken together, these questions about the role that values, interests, and power dynamics play in shaping mission-driven research and its potential impacts evoke critical, though often concealed, interdependencies that exist between the CSS community of practice and the social environments in which its research activities, subject matters, and outputs are embedded. They likewise evoke the inadequacy of evasive scientistic tendencies to appeal to neutral or value-free stances when faced with queries about how values, interests, and power dynamics motivate and influence the aims, purposes, and areas of concern that steer vectors of CSS research. Responding appropriately to such questions surrounding the social determinants of research paths and potential impacts demands an inclusive broadening of the conversations that shape, articulate, and determine the missions to be pursued, the problems to be addressed, and the assessment of potential harms and benefits—a broadening both in terms of the types of knowledge and expertise that are integrated into such deliberative processes and in terms of the range of stakeholder groups that should be involved.

Second, the recognition of a mission-driven and impact-oriented starting point elevates the importance of *identifying the potential adverse effects of CSS research* so that these can, as far as possible, be pinpointed at the outset of research projects and averted. Such practices of anticipatory reflection are necessary because the intended and unintended consequences of the societally impactful insights, tools, and capabilities CSS research produces could be negative and injurious rather than positive and mission-supporting. As the short history of the "big data revolution" demonstrates, the rapid and widespread proliferation of algorithmic systems, data-driven technologies and computation-led analytics has already had numerous deleterious effects on human rights, fundamental freedoms, democratic values, and biospheric sustainability. Such harmful effects have penetrated society at multiple levels including on the planes of individual agency, social interaction, and biospheric integrity. Let us briefly consider these levels in turn.

*Adverse impacts at the individual level*

At the agent level, the predominance "radical behaviourist" attitudes among the academic, industrial, and governmental drivers of data innovation ecosystems have led to the pervasive mobilisation of individual-targeting predictive analytics which have had damaging impacts across a range of human activities (Cardon, 2010; Cohen, 2019b; Zuboff, 2019). For instance, in the domain of e-commerce and ad-tech, strengthening regimes of consumer surveillance have fuelled the use of "large-scale behavioural technologies" (Ball, 2019) that have enabled incessant practices of hyper-personalised psychographic profiling, consumer curation, and behavioural nudging. As critics have observed, such technologies have tended to exploit the emotive vulnerabilities and psychological weaknesses of targeted people (Helbing et al., 2019), instrumentalising them as monetisable sites of "behavioural surplus" (Zuboff, 2020) and treating them as manipulable objects of prediction and "behavioural certainty" rather than as reflective subjects worthy of decision-making autonomy and moral regard (Ball, 2019; Yeung, 2017). Analogous behaviourist postures have spurred state actors and other public bodies to subject their increasingly datafied citizenries to algorithmic nudging techniques that aim to obtain aggregated patterns of desired behaviour which accord with government generated models and predictions (Fourcade & Gordon, 2020; Hern, 2021). Some scholars have characterised such an administrative ambit as promoting the paternalistic displacement of individual agency and the degradation of the conditions needed for the successful exercise of human judgment, moral reasoning, and practical rationality (Fourcade & Gordon, 2020; Spaulding, 2020). In like manner, the nearly ubiquitous scramble to capture behavioural shares of user engagement across online search, entertainment, and social media platforms has led to parallel feedback loops of digital surveillance, algorithmic manipulation, and behavioural engineering (von Otterlo, 2014). The proliferation of the so-called "attention market" business model (Wu, 2019) has prompted digital platforms to measure commercial success in terms of the non-consensual seizure and monopolisation of focused mental activity. This has fostered the deleterious attachment of targeted consumer populations to a growing ecosystem of "distraction technologies" (Syvertsen, 2020; Syvertsen & Enli 2020) and compulsion-forming social networking sites and reputational platforms, consequently engendering, on some accounts, widespread forms of surveillant anxiety (Crawford, 2014), cognitive impairment (Wu 2019), mental health issues (Banjanin et al., 2015; Barry et al., 2017; Lin et al., 2016; Méndez-Diaz et al., 2022; Peterka-Bonetta et al., 2019), and diminished adolescent self-esteem and quality of life (Scott & Woods, 2018; Viner et al., 2019; Woods & Scott, 2016).

*Adverse impacts at the social level*

Setting aside the threats to basic individual dignity and human autonomy that these patterns of instrumentalization, disempowerment, and exploitation present (Aizenberg & van den Hoven, 2020; Halbertal, 2015), the proliferation of data-driven behavioural steering at the collective level has also generated risks to the integrity of social interaction, interpersonal solidarity, and democratic ways of life. In current digital information and communication environments, for example, the predominant steering force of social media and search engine platforms has mobilised opaque computational methods of relevance-ranking, popularity-sorting, and trend-predicting to produce calculated digital publics devoid of any sort of active participatory social or political choice (Gillespie, 2014; Ziewitz, 2016; O'Neil, 2016; Bogost, 2015; Striphas, 2015; Beer, 2017; Cardon, 2016). Rather than being guided by the deliberatively achieved political will of interacting citizens, this vast meshwork of connected digital services shapes these computationally fashioned publics in accordance with the drive to commodify monitored behaviour and to target and capture user attention (Carpentier, 2011; Carpentier & De Cleen 2008; Dean 2010; Fuchs, 2021; John, 2013; Zuckerman, 2020). And, as this manufacturing of digital publics is ever more pressed into the service of profit-seeking by downstream algorithmic mechanisms of hyper-personalised profiling, engagement-driven filtering, and covert behavioural manipulation, democratic agency and participation-centred social cohesion will be increasingly supplanted by insidious forms of social sorting and digital atomisation (Vaidhyanathan 2018; van Dijck, 2013; van Dijck et al., 2018). Combined with complimentary dynamics of wealth polarisation and rising inequality (Wright et al., 2021), such an attenuation of social capital, discursive interaction, and interpersonal solidarity is already underwriting the crisis of social and political polarisation, the widespread kindling of societal distrust, and the animus toward rational debate and consensus-based science that have come to typify contemporary post-truth contexts (Cosentino, 2020; D'Ancona 2017; Harsin, 2018; McIntyre, 2018).

Indeed, as these and similar kinds of computation-based social sorting and management infrastructures continue to multiply, they promise to jeopardise more and more of the formative modes of open interpersonal communication that have enabled the development of crucial relations of mutual trust and responsibility among interacting individuals in modern democratic societies. This is beginning to manifest in the widespread deployment of algorithmic labour and productivity management technologies, where manager-worker and worker-worker relations of reciprocal accountability and interpersonal recognition are being displaced by depersonalising mechanisms of automated assessment, continuous digital surveillance and computation-based behavioural incentivisation, discipline, and control (Ajunwa et al., 2017; Akhtar & Moore, 2016; Kellogg et al., 2020; Moore, 2019). The convergence of the unremitting sensor-based tracking and monitoring of workers' movements, affects, word choices, facial expressions, and other biometric cues, with algorithmic models that purport to detect and correct defective moods, emotions, and levels of psychological engagement and wellbeing may not simply violate a worker's sense of bodily, emotional, and mental integrity by rendering their inner life legible and available for managerial intervention as well as productivity optimisation (Ball, 2009). These forms of ubiquitous personnel tracking and labour management can also have so-called "panoptic effects" (Botan, 1996; Botan & McCreadie, 1990), causing people to alter their behaviour on suspicion it is being constantly observed or analysed and deterring the sorts of open worker-to-worker interactions that enable the development of reciprocal trust, social solidarity, and interpersonal connection. This labour management example merely signals a broader constellation of ethical hazards that are raised by the parallel use of sensor- and location-based surveillance, psychometric and physiognomic profiling (Barrett et al., 2019; Chen & Whitney, 2019; Gifford, 2020; Hoegen et al., 2019; Stark & Hutson, 2021; Agüera y Arcas et al. 2017) and computation-driven technologies of behavioural governance in areas like education (Andrejevic & Selwyn, 2020; Pasquale, 2020), job recruitment (Sanchez-Monedero et al., 2020; Sloane et al, 2022), criminal justice (Brayne, 2020; Pasquale & Cashwell, 2018) and border control (Amoore, 2021; Muller, 2019). The heedless deployment of these kinds of algorithmic systems could have transformative effects on democratic agency, social cohesion, and interpersonal intimacy, preventing people from exercising their freedoms of expression, assembly, and association and violating their right to participate fully and openly in the moral, cultural, and political life of the community.

Lastly, at the level of biospheric integrity and sustainability, the exploding computing power—which has played a major part in ushering in the "big data revolution" and the rise of CSS—has also had significant environmental costs that deserve ethical consideration. As Lannelongue et al. (2021) point out, "the contribution of data centers and high-performance computing facilities to climate change is substantial… with 100 megatonnes of CO2 emissions per year, similar to American commercial aviation". At bottom, this increased energy consumption has hinged on the development of large, computationally intensive algorithmic models that ingest abundant amounts of data in their training and tuning, that undergo iterative model selection and hyperparameter experiments, and that require exponential augmentations in model size and complexity to achieve relatively modest gains in accuracy (Schwartz et al. 2020; Strubell et al., 2019). In real terms, this has meant that the amount of compute needed to train complex, deep learning models increased by 300,000 times in six years (from 2013 to 2019) with training expenditures of energy doubling every six months (Amodei, 2018; Schwartz et al. 2020). Strubell et al. (2019) observe, along these lines, that training Google's large language model, BERT, on GPU, produces substantial carbon emissions "roughly equivalent to a trans-American flight". Though recent improvements in algorithmic techniques, software, and hardware have meant some efficiency gains in the operational energy consumption of computationally hungry, state-of-the-art models, some have stressed that such training costs are increasingly compounded by the carbon emissions generated by hardware manufacturing and infrastructure (e.g. designing and fabricating integrated circuits) (Gupta et al., 2020). Regardless of the sources of emissions, important ethical issues emerge both from the overall contribution of data research and innovation practices to climate change and to the degradation of planetary health and from the differential distribution of the benefits and risks that derive from the design and use of computationally intensive models. As Bender et al. (2021) have emphasised, such allocations of benefits and risks have closely tracked the historical patterns of environmental racism, coloniality, and "slow violence" (Nixon, 2011) that have typified the disproportionate exposure of marginalised communities (especially those who inhabit what has conventionally been referred to as "the Global South") to the pollution and destruction of local ecosystems and to involuntary displacement.

As a whole, these cautionary illustrations of the hazards posed at individual, societal, and environmental levels by ever more ubiquitous computational interventions in the social world should impel CSS researchers to adopt an ethically sober and pre-emptive posture when reflecting on the potential impacts of their projects. The reason for this is not just that many of the methods, tools, capabilities, and epistemic frameworks that they utilise have already operated, in the commercial and political contexts of datafication, as accessories to adverse societal impacts. It is, perhaps more consequentially, that, as Wagner et al. (2021) point out, CSS practices of measurement and corollary theory construction in "algorithmically infused societies… indirectly alter behaviours by informing the development of social theories and subsequently influence the algorithms and technologies that draw on those theories" (p. 197). This dimension of the "performativity" of CSS research—i.e., the way that the activities and theories of CSS researchers can function to reformat, reorganise, and shape the phenomena that they purport only to measure and analyse—is crucial (Healy, 2015; Wagner et al., 2021). It enjoins, for instance, an anticipatory awareness that the methodological predominance of measurement-centred and prediction-driven perspectives in CSS can support the noxious proliferation of the scaled computational manipulation and instrumentalization of large populations of affected people (Eynon et al., 2016; Schroeder, 2014). It also implores cognizance that an unreflective embrace of unbounded sociometrics and the pervasive sensor-based observation and monitoring of research subjects may support wider societal patterns of "surveillance creep" (Lyon, 2003; Marx, 1988) and ultimately have chilling effects on the exercise of fundamental rights and freedoms. The intractable endurance of these kinds of risks of adverse effects and the possibilities for unintended harmful consequences recommends vigilance both in the assessment of the potential impacts of CSS research on affected individuals and communities, and in the dynamic monitoring of the effects of the research outputs, and the affordances they create, once these are released into the social world.

## Challenges related to the quality of CSS research and to its epistemological status

CSS research that is of dubious quality or that misrepresents the world can produce societal harms by misleading people, misdirecting policies, and misguiding further academic research. Many of the pitfalls that can undermine CSS research quality are precipitated by deficiencies in the accuracy and the integrity of the datasets on which it draws. First off, erroneous data linkage can lead to false theories and conclusions. Researchers face ongoing challenges when they endeavour to connect the data generated by identified research subjects to other datasets that are believed to include additional information about those individuals (Weinhardt, 2020). Mismatches can poison downstream inferences in undetectable ways and lead to model brittleness, hampered explanatory power, and distorted world pictures.

The poisoning of inferences by corrupted, inaccurate, invalid, or unreliable datasets can occur in a few other ways. Where CSS researchers are not sufficiently critical of the "ideal user assumption" (Lazer & Radford, 2017), they can overlook instances in which data subjects intentionally mispresent themselves, subsequently perverting the datasets in which they are included. For example, online actors can multiply their identities as "sock puppets" by creating fake accounts that serve different purposes; they can also engage in "gaslighting" or "catfishing" where intentional methods of deception about personal characteristics and misrepresentation of identities are used to fool other users or to game the system; they can additionally impersonate real internet users to purposefully mislead or exploit others (Bu et al., 2013; Ferrara 2015; Lazer & Radford, 2017; Wang et al., 2006; Woolley, 2016; Woolley & Howard, 2018; Zheng et al. 2006). Such techniques of deception can be automated or deployed using various kinds of robots (e.g., chat bots, social media bots, robocalls, spam bots, etc.) (Ferrara et al. 2016; Gupta et al., 2015; Lazer & Radford, 2017; Ott et al. 2011). If researchers are not appropriately attentive to the distortions that may arise in datasets as a result of such non-human sources of misleading data, they can end up unintentionally baking the corresponding corruptions of the underlying distribution that are present in the sample into their models and theories, thereby misrepresenting or painting a false picture of the social world (Ruths & Pfeffer, 2014; Shah et al., 2015). Similar blind spots in detecting dataset corruption can arise when sparse attention is paid to how the algorithms, which pervade the curation and delivery of information on online platforms, affect and shape the data that is generated by the users that they influence and steer (Wagner et al., 2021).

Attentiveness to such data quality and integrity issues can be hindered by the illusion of the veracity of volume or, what has been termed, "big data hubris" (Lazer et al., 2014; Hollingshead et al., 2021; Kitchin, 2015; Mahmoodi et al., 2017). This is the misconception that, in virtue of their sheer volume, big data can "solve all problems", including potential deficiencies in data quality, sampling, and research design (Hollingshead et al., 2021; Meng, 2018). When it is believed that "data quantity is a substitute for knowledge-driven methodologies and theories" (Mahmoodi et al., 2017, p. 57), the rigorous and epistemically vetted approaches to social measurement, theory construction, explanation, and understanding that have evolved over decades in the social sciences and statistics can be perilously neglected or even dismissed.

Such a potential impoverishment of epistemic vigour can also result when CSS researchers fall prey to the enticements of the flip side of big data hubris, namely, computational solutionism. Predispositions to computational solutionism have emerged as a result of the coalescence of the rapid growth of computing power and the accelerating development of complex algorithmic modelling techniques that have together complemented the explosion of voluminous data and the big data revolution. This new access to the computational tools availed by potent compute and high-dimensional algorithmic machinery have led to the misconception in some corners of CSS that tools themselves can, by and large, "solve all problems". Rather than confronting the contextual complexities that lie behind the social processes and historical conditions that generate observational data (Shaw, 2015; Törnberg & Uitermark, 2021), and that concomitantly create manifold possibilities for non-random missingness and meaningful noise, the computational solutionist reverts to a toolbox of heuristic algorithms and technical tricks to "clean up" the data, so that computational analysis can forge ahead frictionlessly (Agniel et al., 2018; Leonelli, 2021). At heart, this contextual sightlessness among some CSS researchers originates in scientistic attitudes that tend to naturalise and reify digital trace data (Törnberg & Uitermark, 2021), treating them as primitive and organically given units of measurement that facilitate the analytical capture of "social physics" (Pentland, 2015), "the 'physics of culture,'" (Manovich, 2016), or the "physics of society" (Caldarelli et al., 2018). The scientific aspiration to discover invariant "laws of society" rests on this erroneous naturalisation of social data. Were the confidence of CSS research in such a naturalist purity of data to be breeched and their contextual and sociohistorical origins appropriately acknowledged, then the scientistic metanarratives that underwrite beliefs in "social physics", and in its nomological character, would consequently be subverted. Computational solutionism provides an epistemic strategy for the wholesale avoidance of this problem: It directs researchers to rely solely on the virtuosity of algorithmic tooling and the computational engineering of observational data to address congenital problems of noise, confounders, and non-random missingness rather than employing a genuine methodological pluralism that takes heed of the critical importance of context and of the complicated social and historical conditions surrounding the generation and construction of data. Such a solutionist tack, however, comes at the cost of potentially misapprehending the circumstantial intricacies and the historically contingent evolution of agential entanglements, social structures, and interpersonal relations and of thereby "misrepresenting the real world" in turn (Ruths & Pfeffer, 2014, p. 1063).

At the epistemic level, this blindered computationalist gambit can also engender social explanations in which the *explanans* (namely, the assemblage of methods, inferences, and evidence used to analyse a phenomenon) does not provide sufficient grounds to account for the *explanandum* (namely, the phenomenon to be explained) (Leslie 2021). An epistemological strategy that attempts to tackle complex sociohistorical problems through quantitatively and computationally anchored methods and explanations alone will simply be insufficient to account for the broad-ranging causal and qualitative factors that can slip through the cracks of quantification, modelling, and simulation, for, as Giglietto et al. (2012) put it, "quantitative [social] data beg for qualitative interpretation" (p.155). On Leonelli's (2021) view, the reductivism of the computationalist strategy can engender a "misguided and autocratic view of science" that excludes "social, contextual factors from evidential reasoning, and thus [disregards] the conditions under which data are generated and interpreted" (Leonelli, 2021). Where epistemological approaches in CSS root out the elucidative role of situated social scientific interpretation that draws on critical insights into the sociohistorical and constructed character of data, they run the risk of neglecting or failing to appropriately represent the contextual nuances of their *explananda*. That is, they risk excluding the kinds of contextually aware reasoning which can holistically weave together a range of explanatory and interpretive understandings to coherently connect *explanantia* to their *explananda* (de Regt et al., 2009).

In addition to these risks posed to the epistemic integrity of CSS by big data hubris and computational solutionism, CSS researchers face another challenge related to the epistemological status of the claims and conclusion they hold forth. This has to do with the problem of interpretability. As the mathematical models employed in CSS research have come to possess ever greater access both to big data and to increasing computing power, their designers have correspondingly been able to enlarge the feature spaces of these computational systems and to turn to gradually more complex mapping functions in order either to forecast future observations or to explain underlying causal structures or effects. In many cases, this has meant vast improvements in the performance of models that have become more accurate and expressive, but this has also meant the growing prevalence of non-linearity, non-monotonicity, and high-dimensional complexity in an expanding array of so-called "black-box" models (Leslie, 2019). Once high-dimensional feature spaces and complex functions are introduced into algorithmic models, the effects of changes in any given input can become so entangled with the values and interactions of other inputs that understanding the rationale behind how individual components are transformed into outputs becomes extremely difficult. The complex and unintuitive curves of many of these models' decision functions preclude linear and monotonic relations between their inputs and outputs. Likewise, the high-dimensionality of their architectures—frequently involving millions of parameters and complex correlations—presents a sweep of compounding statistical associations that range well beyond the limits of human-scale cognition and understanding. Such increasing complexity in input-output mappings creates model opacity and barriers to interpretability. The epistemological problem, here, is that, *as a science that seeks to explain, clarify, and facilitate a better understanding of the human phenomena it investigates*, CSS would seemingly have to avoid or renounce incomprehensible models that obstruct the demonstration of sound scientific reasoning in the conclusions and results attained.

A few epistemic strategies have emerged over the past decade or so to deal with the challenge posed by the problem interpretability in CSS. First, building on a longstanding distinction originally made by statisticians between the predictive and explanatory functions of computational modelling (Breiman, 2001; Mahmoodi et al., 2017; Shmueli, 2010), some CSS scholars have focused on the importance of predictive accuracy, de-prioritising the goals of discovering and explaining the causal mechanisms and reasons that lie behind the dynamics of human behaviour and social systems (Anderson, 2008; Hindman, 2015; Lin, 2015; Yarkoni & Westfall, 2016). Lin (2015), for instance, makes a distinction between the goal of "better science", i.e. "to reveal insights about the human condition"—what Herbert Simon called the "basic science" of explaining phenomena (2002)—and the goal of "better engineering", i.e. "to produce computational artifacts that are more effective according to well-defined metrics" (p. 35)—what Simon called the "applied science" of inferring or predicting from known variables to unknown variables (Simon, 2002; Shmueli, 2010). For Lin, if the purpose of CSS, as an applied science, is "better engineering", then "whatever improves those [predictive] metrics should be exploited without prejudice. Sound scientific reasoning, while helpful, is not necessary to improve engineering". Such a positivistic view would, of course, tamp down or even cast aside the desideratum of interpretability.

However, even for scholars that aspire to retain both the explanatory and predictive dimensions of CSS, the necessity of using interpretable models is far from universally embraced. Illustratively, Hofman et al. (2021) argue for "integrating explanation and prediction in CSS" by treating these approaches as complementary (cf. Engel, 2021; James et al., 2013; Mahmoodi et al., 2017). Still, these authors simultaneously claim that explanatory modelling is about "the estimation of causal effects, regardless of whether those effects are explicitly tied to theoretically motivated mechanisms that are interpretable as 'the cogs and wheels of the causal process'" (Hofman et al., 2021, p. 186). To be sure, they maintain that,

> interpretability is logically independent of both the causal and predictive properties of a model. That is, in principle a model can accurately predict outcomes under interventions or previously unseen circumstances (out of distribution), thereby demonstrating that it captures the relevant causal relationships, and still be resistant to human intuition (for example, quantum mechanics in the 1920s). Conversely, a theory can create the subjective experience of having made sense of many diverse phenomena without being either predictively accurate or demonstrably causal (for example, conspiracy theories) (p. 186-187).

These justifications for treating the goal of interpretability as independent from the causal and predictive characteristics of a model raise some concerns. At an epistemic level, the extreme claim that "interpretability is logically independent of both the causal and predictive properties of a model", is unsupported by the observation that people can be deluded into believing false states of affairs. The attempt to cast aside the principal need for the rational acceptability and justification of the assertoric validity claims that explain a model's causal and predictive properties, because it is possible to be misled by "subjective experience", smacks of a curious epistemological relativism which is inconsistent with the basic requisites of scientific reasoning and deliberation. It offends the "no magic doctrine" (Anderson & Lebiere, 1998) of interpretable modelling, namely, that "it needs to be clear how (good) model performance comes about, that the components of the model are understandable and linked to known processes" (Schultheis, 2022). To level off all adjudications of explanatory claims (strong or weak) about a model because humans can be duped by misled feelings of subjective experience amounts to an absurdity: people can be convinced of bad explanations that are not predictively or causally efficacious (look at all those sorry souls who have fallen prey to conspiracy theories), so all explanations of complex models are logically independent of their actual causal and predictive properties. This line of thinking ends up in a ditch of epistemic whataboutism.

Moreover, at an ethical level, the analogy offered by Hofman et al. between the opaqueness of quantum physics and the opaqueness of "black box" predictive models about human behaviours and social dynamics is misguided and unsupportable. Such an erroneous parallelism is based on a scientistic confusion of the properties of natural scientific variables (like the wave-like mechanics of electrons) that function as heuristics for theory generation, testing, and confirmation in the exact physical sciences and the properties of the social variables of CSS whose generation, construction, and correlation are the result of human choices, evolving cultural patterns, and path dependencies created by sociohistorical structures. Unlike the physics data generated, for instance, by firing a spectroscopic light through a perforated cathode and measuring the splitting of the Balmer lines of a radiated hydrogen spectrum, the all-too-human genealogy of social data means that they can harbour discriminatory biases and patterns of sociohistorical inequity and injustice that become buried within the architectures of complex computational models. In this respect, the "relevant causal relationships" that are inaccessible in opaque models might be fraught with objectionable sociohistorical patterns of inequity, prejudice, coloniality, and structural racism, sexism, ablism, etc. (Leslie et al., 2022a). Because "human data encodes human biases by default" (Packer et al., 2018), complex algorithmic models can house and conceal a troubling range of unfair biases and discriminatory associations—from social biases against gender (Bolukbasi et al., 2016; Lucy & Bamman, 2021; Nozza et al. 2021; Sweeney & Najafian, 2019; Zhao et al., 2017), race (Benjamin, 2019; Noble, 2018; Sweeney, 2013), accented speech (Lawrence, 2021; Najafian et al., 2017), and political views (Cohen & Ruths, 2013; Iyyer et al., 2014; Preoţiuc-Pietro et al. 2017) to structures of encoded prejudice like proxy-based digital redlining (Cottom, 2016; Friedline, 2020) and the perpetuation of harmful stereotyping (Abid et al., 2021; Bommasani et al., 2021; Caliskan et al., 2017; Muñoz et al., 2021; Nadeem et al., 2020; Weidinger et al., 2021). A lack of interpretability in complex computational models whose performant causal and predictive properties could draw opaquely on secreted discriminatory biases or patterns of inequity is therefore ethically intolerable. As Wallach (2018) observes,

> the use of black box predictive models in social contexts…[raises] a great deal of concern—and rightly so—that these models will reinforce existing structural biases and marginalize historically disadvantaged populations… we must [therefore] treat machine learning for social science very differently from the way we treat machine learning for, say, handwriting recognition or playing chess. We cannot just apply machine learning methods in a black-box fashion, as if computational social science were simply computer science plus social data. We need transparency. We need to prioritize interpretability—even in predictive contexts" (p.44)(cf. Lazer et al., 2020, p. 1062).

## Challenges related to research integrity

Challenges related to research integrity are rooted in the asymmetrical dynamics of resourcing and influence that can emerge from power imbalances between the CSS research community and the corporations and government agencies upon whom CSS scholars often rely for access to the data resources, compute infrastructures, project funding opportunities, and career advancement prospects they need for their professional subsistence and advancement. Such challenges can manifest, inter alia, in the exercise of research agenda-setting power by private corporations and governmental institutions, which set the terms of project funding schemes and data sharing agreements, and in the willingness of CSS researchers to produce insights and tools that support scaled behavioural manipulation and surveillance infrastructures.

These threats to the integrity of CSS research activity manifests in a cluster of potentially unseemly alignments and conflicts of interest between its own community of practice and those platforms, corporations, and public bodies who control access to the data resources and compute infrastructures upon which CSS researchers depend (Theocharis & Jungherr, 2021). First, there is the potentially unseemly alignment between the extractive motives of digital platforms, which monetise, monger, and link their vast troves of personal data and marshal inferences derived from these to classify, mould, and behaviourally nudge targeted data subjects, and the professional motivations CSS researchers who desire to gain access to as much of this kind of social big data as possible (Törnberg & Uitermark, 2021). A similar alignment can be seen between the motivations of CSS researchers to accumulate data and the security and control motivations of political bodies, which collect large amounts of personal data from the provision and administration of essential social goods and services often in the service of such motivations (Fourcade & Gordon, 2020). There is also a potentially unseemly alignment between the epistemic leverage and sociotechnical capabilities desired by private corporations and political bodies interested in scaled behavioural control and manipulation and the epistemic leverage and sociotechnical capabilities cultivated, as a vocational *raison d'être*, by some CSS researchers who build predictive tools. This alignment is made all-the-more worrying by the asymmetrical power dynamics that can be exercised by the former organisations over the latter researchers, who not only are increasingly reliant on private companies and governmental bodies for essential data access and computing resources but are also increasingly the obliged beneficiaries of academic-corporate research partnerships and academic-corporate 'dual-affiliation' career trajectories that are funded by large tech corporations (Roberge et al., 2019). Finally, there is a broader scale cultural alignment between the way that digital platforms and tech companies pursue their corporate interests through technology practices that privilege considerations of strategic control, market creation, and efficiency and that are thereby functionally liberated from the constraints of social licence, democratic governance, and considerations of the interests of impacted people (Feenberg, 2002, 2012) and the way that CSS scholars can pursue of their professional interests through research practices similarly treated as operationally autonomous and independent from the societal conditions they impact and the governance claims of affected individuals and communities.

## Challenges related to research equity

Challenges related to research equity fall under two categories: (1) Inequities that arise within the outputs of CSS research in virtue of biases that crop up within its methods and analytical approaches, and (2) Inequities that arise within the wider field of CSS research that result from material inequalities and capacity imbalances between different research communities. Challenges emerging from the first category include the potential reinforcement of digital divides and data inequities through biased sampling techniques that render digitally marginalised groups invisible as well as potential aggregation biases in research results that mask meaningful differences between studied subgroups and therefore hide the existence of real-world inequities. Challenges emerging from the second category include exploitative data appropriation by well-resourced researchers and the perpetuation of capacity divides between research communities, both of which derive from long-standing dynamics of regional and global inequality that may undermine reciprocal sharing and collaboration between researchers from more and less resourced geographical areas, universities, or communities of practice.

Issues of sampling or population bias in CSS datasets extracted from social media platforms, internet use, and connected devices arise when the sampled population that is being studied differs from the larger target population in virtue of the non-random selection of certain groups into the sample (Mehrabi et al., 2021; Hargittai, 2015, 2020; Hollingshead et al., 2021; Olteanu et al., 2019; Tufekci, 2014). It has been widely observed that people do not select randomly into social media sites like Twitter (Blank, 2017; Blank & Lutz, 2017), MySpace (boyd, 2011), Facebook (boyd, 2011; Hargittai, 2015), and LinkedIn (Blank & Lutz, 2017; Hargittai, 2015). As Hargittai (2015) shows, in the US context, people with greater educational attainment and higher income were more likely to be users of Twitter, Facebook, and LinkedIn than others of less privilege. Hargittai (2020) claims, more generally, that "big data derived from social media tend to oversample the views of more privileged people" and people who possess greater levels of "internet skill". Earlier studies and surveys have also demonstrated that, at any given time, "different user demographics tend to be drawn to different social platforms" (Olteanu et al., 2019), with men and urban populations significantly over-represented among Twitter users (Mislove et al., 2011) and women over-represented on Pinterest (Ottoni et al., 2013).

The oversampling of self-selecting privileged and dominant groups, and the under-sampling or exclusion of members of other groups who may lack technical proficiency, digital resources, or access to connectivity, for example, large portions of elderly populations (Friemel, 2016; Haight, Quan-Haase, & Corbett, 2014; Quan-Haase, Williams, Kicevski, Elueze, & Wellman, 2018), can lead to an inequitable lack of representativity in CSS datasets—rendering those who have been left out of data collection for reason of accessibility, skills, and resource barriers "digitally invisible" (Longo et al., 2017). Such sampling biases can cause deficiencies in the ecological validity of research claims (Olteanu et al., 2019), impaired performance of predictive models for non-majority subpopulations (Johnson et al., 2017), and, more broadly speaking, the failure of CSS models to generalise from sampled behaviours and opinions to the wider population (Blank, 2017; Hargittai & Litt, 2012; Hollingshead, 2021). This hampered generalisability can be especially damaging when the insights and results of CSS models, which oversample privileged subpopulations and thus disadvantage those missing from datasets, are applied willy-nilly to society as a whole and used to shape the policymaking approaches to solving real-world problems. As Hollingshead et al. (2021) put it, "the ethical concern here is that, as policymakers and corporate stakeholders continue to draw insights from big data, the world will be recursively fashioned into a space that reflects the material interests of the infinitely connected" (p. 173).[7]

---

[7] A similar and compounding form of sampling bias can occur when survey data is linked, through participant consent, to digital trace data from social media networks. Here the dynamic of non-random self-selection manifests

Another research inequity that can crop up within CSS methods and analytical approaches is aggregation bias (Mehrabi et al., 2021; Suresh & Guttag, 2021). This occurs when a model's analysis is applied in a "one-size-fits-all" manner to subpopulations that have different conditional distributions, thereby treating the results as "population-level trends" that map inputs to outputs uniformly across groups despite their possession of diverging characteristics (Hollingshead et al., 2021; Suresh & Guttag, 2019). Such aggregation biases can lead models to fit optimally for dominant or privileged subpopulations that are oversampled while underperforming for groups that lack adequate representation. These biases can also conceal patterns of inequity and discrimination that are differentially distributed among subpopulations (boyd & Crawford, 2012; Barocas & Selbst, 2016; Hollinghead, 2022; Longo et al., 2017; Olteanu et al., 2019), consequently entrenching or even augmenting structural injustices that are hidden from view on account of the irresponsible statistical homogenisation of target populations.

A different set of research inequities arise within the wider field of CSS research as a consequence of material inequalities and capacity imbalances that exist between different research communities. Long-standing dynamics of global inequality, for instance, may undermine reciprocal sharing between research collaborators from high-income countries (HICs) and those from low-/middle-income countries (LMICs) (Leslie, 2020). Given asymmetries in resources, infrastructure, and research capabilities, data sharing between LMICs and HICs, and transnational research collaboration, can lead to inequity and exploitation (Bezuidenhout et al., 2017; Leonelli, 2013; Shrum, 2005). That is, data originators from LMICs may put immense amounts of effort and time into developing useful datasets (and openly share them) only to have their countries excluded from the benefits derived by researchers from HICs who have capitalized on such data in virtue of greater access to digital resources and compute infrastructure (Goldacre et al., 2015). Moreover, data originators from LMICs may generate valuable datasets that they are then unable to independently and expeditiously utilize for needed research, because they lack the aptitudes possessed by researchers from HICs who are the beneficiaries of arbitrary asymmetries in education, training, and research capacitation (Bull et al., 2015; Merson et al., 2015).

This can create a two-fold architecture of research inequity wherein the benefits of data production and sharing do not accrue to originating researchers and research subjects, and the scientists from LMICs are put in a position of relative disadvantage vis-à-vis those from HICs whose research efficacy and ability to more rapidly convert data into insights function, in fact, to undermine the efforts of their disadvantaged research partners (Bezuidenhout et al., 2017; Crane, 2011). It is important to note, here, that such gaps in research resources and capabilities also exist within HICs where large research universities and technology corporations (as opposed to less well-resourced universities and companies) are well positioned to advance data research given their access to data and compute infrastructures (Ahmed & Wahed, 2020).

In redressing these access barriers, emphasis must be placed on "the social and material conditions under which data can be made useable, and the multiplicity of conversion factors required for researchers to engage with data" (Bezuidenhout et al., 2017, p. 473). Equalizing know-how and capability is a vital counterpart to equalizing access to resources, and both together are necessary preconditions of just research environments. CSS scholars engaging in international research collaborations should focus on forming substantively reciprocal partnerships where capacity-building and asymmetry-aware practices of cooperative innovation enable participatory parity and thus greater research access and equity.

# Incorporating habits of responsible research and innovation into CSS practices

The foregoing taxonomy of the five main ethical challenges faced by CSS is intended to provide CSS researchers with a critical lens that enables them to sharpen their field of vision so that they are equipped

---

in the select group of research subjects (likely those who are privileged and young and more frequently male) who have social media accounts and who consent to having them linked to the survey research (Al Baghal et al., 2019; Stier, Breuer, Siegers, & Thorson, 2019).

to engage in the sort of anticipatory reflection which roots out irresponsible research practices and harmful impacts. However, circumvention of the potential endurance of "research fast and break things" attitudes requires a deeper cultural transformation in the CSS community of practice. It requires the end-to-end incorporation of habits of Responsible Research and Innovation (RRI) into all its research activities. An RRI perspective provides CSS researchers with an awareness that all processes of scientific discovery and problem-solving possess sociotechnical aspects and ethical stakes. Rather than conceiving research as independent from human values, RRI regards these activities as ethically-implicated social practices. For this reason, such practices are charged with a responsibility for *critical self-reflection* about the role that these values play both in discovery, engineering, and design processes and in considerations of the real-world effects of the insights and technologies that these processes yield.

Those who have been writing on the ethical dimension of CSS for the past decade have emphasised the importance of precisely these kinds of self-reflective research practices (for instance, BSA, 2016; Eynon & Schroeder, 2016; franzke et al., 2020; Hollingshead et al., 2021; Lomborg, 2013; Markham & Buchanan, 2012; Moreno et al., 2013; Weinhardt, 2020). Reacting to recent miscarriages of research ethics that have undermined public trust, such as the 2016 mass sharing of sensitive personal information that had been extracted from researchers from the OKCupid dating site (Zimmer, 2016), they have stressed the need for "a bottom-up, case-based approach to research ethics, one that emphasizes that ethical judgment must be based on a sensible examination of the unique object and circumstances of a study, its research questions, the data involved, and the expected analysis and reporting of results, along with the possible ethical dilemmas arising from the case" (Lomborg, 2013, p. 20). What is needed to operationalise such a "a bottom-up, case-based approach to research ethics" is the development across the CSS community of habits of RRI. In this section, we will explore how CSS practices can incorporate habits of RRI, focusing, in particular, on the role that contextual considerations, anticipatory reflection, public engagement, and justifiable action should play across the research lifecycle.

Building on research in Science and Technology Studies and Applied Technology Ethics, the RRI view of 'science with and for society' has been transformed into helpful general guidance in such interventions as EPSRC's 2013 AREA framework and the 2014 Rome Declaration (Fisher & Rip 2013; Owen, 2014; Owen et al. 2013; Owen et al., 2012; Stilgoe et al. 2013; von Schomberg 2013). More recently, EPSRC's AREA principles (Anticipate, Reflect, Engage, Act) have been extended into the fields of data science and AI by the CARE & Act Framework (Consider context, Anticipate impacts, Reflect on purposes, positionality, and power, Engage inclusively, Act responsibly and transparently) (Leslie, 2020; Leslie et al., 2022c). The application of the CARE & Act principles to CSS aims to provide a handy tool that enables its researchers to continuously sense check the social and ethical implications of their research practices and that helps them to establish and sustain responsible habits of scientific investigation and reporting. Putting the CARE & Act Framework into practice involves taking its several guiding maxims as a launching pad for continuously reflective and deliberate choice-making across the research workflow. Let us explore each of these maxims in turn.

## Consider context

The imperative of considering context enjoins CSS researchers to think diligently about the conditions and circumstances surrounding their research activities and outputs. This involves focusing on the norms, values, and interests that inform the people undertaking the research and that shape and motivate the reasonable expectations of research subject and those who are likely to be impacted by the research and its results: How are these norms, values and interests influencing or steering the project and its outputs? How could they influence research subjects' meaningful consent and expectations of privacy, confidentiality, and anonymity? How could they shape a research project's reception and impacts across impacted communities? Considering context also involves taking into account the specific domain(s), geographical location(s), and jurisdiction(s) in which the research is situated and reflecting on the expectations of affected stakeholders that derive these specific contexts: How are do the existing institutional norms and rules in a given domain or jurisdiction shape expectations regarding research goals, practices, and outputs? How do the the unique social, cultural, legal, economic, and political environments in which different research projects are embedded influence the conditions of data generation, the intentions and behaviours of the research subjects that are captured by extracted data, and the space of possible inferences that data analytics, modelling, and simulation can yield?

The importance of responsiveness to context has been identified as significant in internet research ethics for nearly two decades (Buchanan, 2011; Markham, 2006) and has especially been emphasised more recently in the *Internet Research: Ethical Guidelines 3.0* of the Association of Internet Researchers (AoIR), where the authors stress that a "basic ethical approach" involves focussing on "on the fine-grained contexts and distinctive details of each specific ethical challenge" (franzke, 2020, p. 4).[8] For franzke et al., such a

> process- and context-oriented approach… helps counter a common presumption of "ethics" as something of a 'one-off' tick-box exercise that is primarily an obstacle to research. On the contrary…taking on board an ongoing attention to ethics as inextricably interwoven with method often leads to better research as this attention entails improvements on both research design and its ethical dimensions throughout the course of a project" (pp. 4-5).

This ongoing attention entails a keen awareness of the need to "respect people's values or expectations in different settings" (Eynon et al. 2016) as well as the need to acknowledge cultural differences, ethical pluralism, and diverging interpretations of moral values and concepts (Capurro, 2005, 2008; Ess, 2020; Hongladarom and Ess, 2007; Leslie et al., 2022a). Likewise, contextual considerations need to include a recognition of interjurisdictional differences in legal and regulatory requirements (for instance, variations in data protection laws and legal privacy protections across regions and countries whence digital trace data is collected).

All in all, contextual considerations should, at minimum, track three vectors: The first involves considering the contextual determinants of the condition of the production of the research (e.g., thinking about the positionality of the research team, the expectations of the relevant CSS community of practice, and the external influences on the aims and means of research by funders, collaborators, and providers of data and research infrastructure); The second involves considering the context of the subjects of research (e.g., thinking about research subjects' reasonable expectations of gainful obscurity and 'privacy in public' and considering the changing contexts of their communications such as with whom they are interacting, where, how, and what kinds of data are being shared); The third involves considering the contexts of the social, cultural, legal, economic, and political environments in which different research projects are embedded as well as the historical, geographic, sectoral, and jurisdictional specificities that configure such environments (e.g., thinking about the ways different social groups—both within and between cultures—understand and define key values, research variables, and studied concepts differently as well as the ways that these divergent understandings place limitations on what computational approaches to prediction, classification, modelling, and simulation can achieve).

## Anticipate impacts

The imperative of anticipating impacts enjoins CSS researchers to reflect on and assess the potential short-term and long-term effects their research may have on impacted individuals (e.g., research participants, data subjects, and the researchers themselves) and on affected communities and social groups, more broadly. The purpose of this kind of anticipatory reflection is *to safeguard the sustainability of CSS projects across the entire research lifecycle*. To ensure that the activities and outputs of CSS research remain socially and environmentally sustainable and support the sustainability of the communities they affect, researchers must proceed with a continuous responsiveness to the real-world impacts that their research could have. This entails concerted and stakeholder-involving exploration of the possible adverse and beneficial effects that could otherwise remain hidden from view if deliberate and structured processes for anticipating downstream impacts were not in place. Attending to sustainability, along these lines, also entails the iterative re-visitation and re-evaluation of impact assessments. To be sure, in its general usage, the word "sustainability" refers to the maintenance of and care for an object or endeavour *over time*. In the CSS context, this implies that building sustainability into a research project is not a "one-off" affair. Rather, carrying out an initial research impact assessment at the inception of a project is only a first, albeit critical, step in a much longer, end-to-end process of responsive re-evaluation and re-assessment. Such an iterative approach enables sustainability-aware researchers to pay continuous attention both to the dynamic and changing character of the research lifecycle and to the shifting conditions of the real-world environments in which studies are embedded.

---

[8] It is important to note that the importance of contextual considerations has also been present in earlier versions of the AoIR guidelines which date back two decades (IRE 1.0, 2002; IRE 2.0, 2012).

This demand to anticipate research impacts is not new in the modern academy—especially in the biomedical and social sciences, where IRB processes for research involving human subjects have been in place for decades (Abbott and Grady, 2011; Grady 2015). However, the novel human scale, breadth, and reach of CSS research, as well as the new (and often subtler) range of potential harms it poses to impacted individuals, communities, and the biosphere, call into question the adequacy of conventional IRB processes (Metcalf and Crawford, 2016). While the latter have been praised as a necessary step forward in protecting the physical, mental, and moral integrity of human research subjects, building public trust in science, and institutionalising needed mechanisms for ethical oversight (Resnik, 2018), critics have also highlighted their unreliability, superficiality, narrowness, and inapplicability to the new set of information hazards posed by the processing of aggregated big data (Prunkl et al., 2021; Raymond, 2019).

A growing awareness of these deficiencies has generated an expanding interest in CSS-adjacent computational disciplines (like machine learning, artificial intelligence, and computational linguistics) to come up with more robust impact assessment regimes and ethics review processes (Hecht et al., 2021; Leins et al., 2020; Nanayakkara, 2021). For instance, in 2020, the NeurIPS conference introduced a new ethics review protocol that required paper submissions to include an impact statement "discussing the broader impact of their work, including possible societal consequences—both positive and negative" (NeurIPS, 2020). Informatively, this protocol was converted into a responsible research practices checklist in 2021 (NeurIPS, 2021) after technically oriented researchers protested that they lacked the training and guidance needed to carry out impact assessments effectively (Ashurst et al., 2021; Johnson, 2020; Prunkl et al., 2021). Though there has been recent progress made, in both AI and CSS research communities, to integrate some form of ethics training into professional development (Ashurst et al., 2020; Salganik and SICSS, nd.) and to articulate guidelines for anticipating ethical impacts (NeurIPS, 2022), there remains a lack of institutionalised instruction, codified guidance, and professional stewardship for research impact assessment processes. As an example, conferences such as International AAAI Conference on Web and Social Media - ICWSM (2022), International Conference on Machine Learning - ICML (2022), North American Chapter of the Association for Computational Linguistics - NAACL (2022), and Empirical Methods in Natural Language Processing – EMNLP (2022) each require some form of research impact evaluation and ethical consideration, but aside from directing researchers to relevant professional guidelines and codes of conduct (e.g., from the Association for Computational Linguistics - ACL, Association for Computing Machinery - ACM, and Association for the Advancement of Artificial Intelligence - AAAI), there is scant direction on how to operationalise impact assessment processes (Prunkl et al., 2021).

What is missing from this patchwork of ethics review requirements and guidance is a set of widely accepted procedural mechanisms that would enable and standardise conscientious research impact assessment practices. To fill this gap, recent research into the governance practices needed to create responsible data research environments has called for a coherent, integrated, and holistic approach to impact assessment that includes several interrelated elements (Leslie 2019, 2020; Leslie et al., 2021, 2022b, 2022c, 2022d):

*Stakeholder analysis*: Diligent research impact assessment practices should include processes that allows researchers to identify and evaluate the salience and contextual characteristics of individuals or groups who may be affected by, or may affect, the research project under consideration (Mitchell et al., 2017; Reed et al., 2005; Schmeer, 1999; Varvasovszky and Brugha, 2000). Stakeholder analysis aims to help researchers understand the relevance of each identified stakeholder to their project and to its use contexts.[9] It does this

---

[9] Scholars and practitioners from areas as diverse as public policy, land use, environmental and natural resource management, international development, and public health have offered many different definitions of "stakeholders" over the past several decades. Even so, these definitions have converged around a few common characteristics. Stakeholders are individuals or groups that (1) have interests or rights that may be affected by the past, present, and future decisions and activities of an organisations; (2) may have the power or authority to

by providing a structured way to assess the relative interests, rights, vulnerabilities, and advantages of identified stakeholders as these characteristics may be impacted by, or may impact, the research.

Three steps are involved in thorough stakeholder analysis. First, researchers should draw on desk-based research, domain expertise, local knowledge, and the lived experience of relevant community members to get a sense of the social environment and human factors that may be affected by, or may affect, the research. This initial exploration should also include positionality reflection to help determine whether the backgrounds of researchers could introduce biases or blind spots into the analysis (elaborated on in the next section). Second, building on this contextual understanding, researchers should identify those individuals and groups who may be significantly impacted by, or may impact, the project, paying close attention to vulnerable and protected groups. Finally, researchers should carry out a stakeholder salience analysis to determine the individuals and groups who are most relevant when considering potential project impacts. This involves assessing the relative interests, rights, vulnerabilities, and advantages of identified stakeholders as these characteristics may be impacted by the project.

Stakeholder analyses may be carried out in a variety of ways that involve more-or-less stakeholder involvement. This spectrum of options ranges from analyses carried out exclusively by a research team without active community engagement to analyses built around the inclusion of community-led participation and co-design from the earliest stages of stakeholder identification. The degree of stakeholder involvement should vary from project to project based upon a preliminary assessment of the potential risks and hazards of the research, with stakeholder engagement being proportionate to the severity and scale of the potential dangers posed by the project.

*Establishment of clear normative criteria for impact assessment*: Effective research impact assessment practices should start from a clear set of ethical values or human rights criteria against which the potential impacts of a project on affected individuals and communities can be evaluated. Such criteria should provide common but non-exclusive point of departure for collective deliberation about the ethical permissibility of the research project under consideration. Adopting common normative criteria from the outset enables reciprocally respectful, sincere, and open discussion about the ethical challenges a research project may face by helping to create a shared vocabulary for informed dialogue and impact assessment. Such a common starting point also facilitates deliberation about how to balance ethical values when they come into tension.

There is, however, a crucial hurdle that must be cleared when establishing which normative criteria to adopt. Amid the undeniable ethical plurality of modern social life, it has become essential to acknowledge the *historically relative* and *contextually situated* character of normative criteria *per se* (Ess, 2006; Lassman, 2011; Madsen and Strong, 2009). This implies that no fixed or universally accepted list of ethical values or fundamental rights and freedoms could *pre-reflectively* provide such a common starting point. Over the past several decades, research ethicists have, for this reason, taken a more pragmatic and empirically driven position, in proposing basic values, that begins by considering the set of real-world dangers posed by practices of scientific research and by the use of the innovations they yield. Indeed, the principles that have emerged from the two main sources of modern Western research ethics, namely, bioethics and human rights, have found their origins in moral claims that have responded directly to tangible, technologically inflicted harms and atrocities. Whereas the human rights perspective (and its expressions in the founding documents of research ethics, the 1947 Nuremberg Code and the 1964 Helsinki Declaration) has its roots in efforts to redress the well-known technological barbarisms and genocides of the mid-twentieth century, in the case of bioethics, its emergence tracked the public exposure in the 1960s and 1970s of several atrocities of human experimentation—such as the infamous Tuskegee syphilis experiment in the US.[10]

---

influence the outcome of such decisions and activities; (3) possess relevant characteristics that put them in positions of advantage or vulnerability with regard to those decisions and activities.

[10] In instances like the Tuskegee syphilis experiment, it was discovered that members of vulnerable or marginalized social groups had been subjected to the injurious effects of institutionally run biomedical experiments without having knowledge of or giving consent to their participation.

The responsiveness of the principles of human rights and bioethics to technological harms goes some way to explaining their prominence in contemporary digital ethics. Across all the iterations of the Internet Research Ethics guidelines (IRE 1.0, IRE 2.0, and IRE 3.0), the "Primary Ethical Norms", which are taken as basic normative criteria (respect for persons, beneficence, and justice), are drawn directly from bioethics (Beauchamp, 2008; franzke et al., 2020). Likewise, in the applied ethics of artificial intelligence and data science, ethics researchers have broadly converged around human rights and bioethical principles that are seen as effectually responding to the real-world problems posed by the use of the AI and data-driven technologies themselves. These hazards include the potential loss of human agency, privacy, and social connection in the wake of expanding automation and datafication, harmful outcomes that may result from the use of poor-quality data or poorly designed systems, and the possibility that entrenched societal dynamics of bias and discrimination will be perpetuated or even augmented by data-driven technologies that tend to reinforce existing social and historical patterns. Accordingly, principles like protecting human dignity, respecting the integrity of private and family life, ensuring solidarity and social connection, supporting human and biospheric wellbeing, and safeguarding equal status, social justice, and the common good have emerged as widely accepted normative criteria (Council of Europe, 2020; High Level Expert Group on AI, 2019; Institute Of Electrical And Electronics Engineers, 2018; Leslie, 2019; Toronto Declaration, 2018; University of Montreal, 2017).

While these ethical values provide a solid basis for research impact assessment in CSS, there is another valence of ethical plurality that must be confronted. From a more interculturally oriented perspective, researchers must acknowledge that the exclusion of non-Western ethical frameworks from the dominant discourses that have shaped the ethics and governance of digital technologies and computational research up to the present reflects deeper legacies of coloniality and Western cultural hegemony that are in need of redress (de Sousa Santos, 2018 Medina, 2012; Quijano, 2007). On this view, given the planetary stretch of CSS research, its research ethics must confront the way that such legacies have created an unsustainable homogeneity of ethical values in digital ethics. Resistance to the prevailing the monoculture of Western-centric morality will allow CSS research ethics to become sufficiently responsive to the condition of cultural and ethical pluralism that typifies the modern, interconnected global society both between nation-states and regions and within them (Aggarwal, 2020; franzke et al., 2020; Leslie et al., 2022a).[11]

Any normative criteria that form the basis for research impact assessment must thus be inclusive of the diverse cultural self-understandings and lived experience of all those who may be affected whether or not their value standpoints lie within predominant Western sociocultural sensibilities (Birhane, 2021; Mhlambi, 2020). To meet such a need for an ethically pluralistic and normatively inclusive approach to CSS research ethics, the establishment of clear normative criteria for impact assessment must reflect and foster non-Western visions of ethical life—visions that often depart from the predominant individualistic ethos of Anglo-European framings and instead embrace a more relational, biocentric, and community-based view of moral action and interaction (such as seen, for instance, in the Ubuntu affirmation of moral personhood through social relationality or in the Abya Yala Indigenous prioritization of living well, *sumac kawsay*, and the care for Mother Earth, *Pachamama,* in South America) (Eze, 2008; Gyekye, 1992; Huanacuni 2010; Kalumba, 2020; Mbiti, 1970; Menkiti, 1984; Walsh 2015, 2018). Attempts to actualise this broadened scope of normative criteria for assessment of the impacts of computational research and innovation have been recently made in UNESCO's "Recommendation on the ethics of artificial intelligence", which has been adopted by its 193 member states, and in the "12 Principles and Priorities of Responsible Data Innovation" proposed as part of the Global Partnership on AI's (GPAI's) 2021-2022 *Advancing Data Justice Research and Practice* project (include here as Annex 1).

*Methodical evaluation of potential impacts and impact mitigation planning*: The actual research impact assessment process provides an opportunity for research teams (and engaged stakeholders, where deemed appropriate) to produce detailed evaluations of the potential and actual impacts that the project may have, to contextualize and corroborate potential harms and benefits, to make possible the collaborative assessment of the severity of potential adverse impacts identified, and to facilitate the co-design of an impact mitigation plan.

---

[11] Notwithstanding the importance of this critical trajectory, as Chakrabarty (2000) emphasises, such a corrective view does not entail a wholesale rejection or dismissal of the Western intellectual traditions that have given rise to the "universal and secular vision of the human" in which key components of human rights and fundamental freedoms are anchored. The latter vision, as Chakrabarty argues, has provided indispensable normative leverage for the pursuit of justice, the battle against oppression and inequity, and the advancement of human freedom in both national and post-colonial/de-colonial contexts.

As noted above, methodical impact evaluation should involve an initial adoption of normative criteria that function as metrics for scoping and assessing the possible harms and benefits of the research and its outputs. Taking GPAI's "12 Principles and Priorities of Responsible Data Innovation" (Annex1) as an example, relevant impact assessment questions could include:

*How, if at all, could our research and its outputs impact each of the following twelve principles and priorities as they relate to all affected stakeholders, especially those who are vulnerable, marginalised, or historically discriminated against? (Affected stakeholders include research subjects and participants, subjects of data collected for or used in the study, researchers, and all other impacted people and social groups.)*

- *Respect for and protection of human dignity*
- *Interconnectivity, solidarity, and intergenerational reciprocity*
- *Environmental flourishing, sustainability, and the rights of the biosphere*
- *Protection of human freedom and autonomy*
- *Prevention of harm and protection of the right to life and physical, psychological, and moral integrity*
- *Non-discrimination, fairness, and equality*
- *Rights of Indigenous peoples and Indigenous data sovereignty*
- *Data protection and the right to respect of private and family life*
- *Economic and social rights*
- *Accountability and effective remedy*
- *Democracy*
- *Rule of law*

*How could our research and its outputs advance each of these twelve principles and priorities or hinder their realisation?*

*Are there particular stakeholder groups who could disproportionately enjoy the benefits of the research and its outputs, or suffer from the potential harms they generate, as these harms and benefits relate to each of the twelve principles and priorities?*

*If things go wrong in our research or if its outputs (especially tools produced or capacities enabled) are used out-of-the-scope of their intended purpose and function, what harms could be done to stakeholders in relation to each of the twelve principles and priorities?*

It is important to note here that stakeholder involvement in the research impact assessment process can be a critical safeguard against evaluative blind spots and omissions. Impacted individuals and social groups are often in a better position to identify salient impacts, and the inclusion of affected people in impact evaluation processes enables research teams to appropriately contextualize and corroborate the potential harms and benefits they discern in dialogue with people whose positionality and lived experience well situates them to reflectively anticipate possible hazards and advantages.

Methodical impact evaluation should also involve an assessment of the severity of potential adverse impacts. This brings clarity to the prioritisation of impact mitigation actions by allowing the severity levels of potential negative effects to be differentiated, elucidated, and refined. As explained in the United Nations Guiding Principles on Business and Human Rights (UNGP), assessing the severity of potential negative impacts on fundamental rights and freedoms involves consideration of their scale, scope, and remediability, where scale is defined as "the gravity or seriousness of the impact," scope as "how widespread the impact is, or the numbers of people impacted," and remediability, as the "ability to restore those affected to a situation at least the same as, or equivalent to, their situation before the impact" (UNGP, 2011, Principle 14).[12]

One notable challenge faced by researchers who are assessing the severity of potential adverse impacts is *identifying cumulative or aggregate impacts of the research and its outputs* on stakeholders (and their progeny) that could expand their effects beyond the scope of impact identified for those individuals and communities who are directed affected. Identifying cumulative or aggregate downstream impacts can be much more difficult than identifying harms directly or proximately caused by a research project and its outputs, and discerning these impacts may require additional research and consultation with domain experts and other relevant stakeholders. This difficulty results from the fact that cumulative impacts are often incremental and more difficult to perceive, and they frequently involve complex contexts of multiple actors or research projects operating in the same area or sector or affecting the same populations (Götzmann et al., 2020). Some "big picture" questions to reflect on when assessing cumulative or aggregate impacts include:

> *Could the research and its outputs contribute to wider scale adverse impacts when its deployment is coordinated with (or occurs in tandem with) other research projects or innovation activity that serve similar functions or purposes? For example, if the impacts of a CSS project that aims to discover an effective method of behavioural nudging at scale are considered in combination with the proliferation of many other similar projects or computational systems in a given sector, concerns about wider cumulative effects like mass manipulation, objectification, and infringement on autonomy and human dignity become relevant.*

> *Could the research and its outputs replicate, reinforce, or augment socio-historically entrenched legacy harms that create knock-on effects in impacted individuals and groups? For example, if a CSS project analyses sensitive personal information contained in databases scraped from social media websites without gaining the proper consent of research subject in accordance with their reasonable expectations, it could add to the legacy harms of companies that have used data recklessly and eroded public trust regarding the respect of privacy and data protection rights in the digital sphere. This can create wider chilling effects on elements of open communication, information sharing, and interpersonal connection that are essential components for the sustainability of democratic forms of life.*

> *Could the production and use of the system be understood to contribute to wider aggregate adverse impacts on the biosphere and on planetary health when its deployment is considered in combination with other systems that may have similar environmental impacts? For example, a CSS project that involves moderate levels of energy consumption in model training or data storage may be seen to contribute to significant environmental impact when considered alongside the energy consumption of similar projects across research ecosystems.*

Once impacts have been evaluated and the severity of any potential harms assessed, impact prevention and mitigation planning should commence. Diligent impact mitigation planning begins with a scoping and prioritization stage. Research team members (and engaged stakeholders, where appropriate) should go through all the identified potential adverse impacts and map out the interrelations and interdependencies between them as well as surrounding social factors (such as contextually specific stakeholder vulnerabilities and precariousness) that could make impact mitigation more challenging. Where prioritization of prevention and mitigation actions is necessary (for instance, where delays in addressing a potential harm could reduce its remediability), decision-making should be steered by the relative severity of the impacts under consideration. As a general rule, while impact prevention and mitigation planning may involve prioritization of actions, all potential adverse impacts must be addressed. When potential adverse impacts have been mapped out and organised, and mitigation actions have been considered, the research team (and engaged stakeholders, where appropriate) should begin co-designing an impact mitigation plan (IMP). The IMP will become the part of your transparent reporting methodology that specifies the actions and processes needed to address the adverse impacts which have been identified and that assigns responsibility for the completions of these tasks and processes. As such, the IMP will serve a crucial documenting function.

---

[12] For detailed guidance on assessing the scale, scope, and remediability aspects of severity, see (Leslie et al. 2021, pp. 238-246).

*Establishment of protocols for re-visitation and re-evaluation of the research impact assessment*: Research impact assessments must pay continuous attention both to the dynamic and changing character of the research lifecycles and to the shifting conditions of the real-world environments in which research practices, results, and outputs are embedded. There are two sets of factors that should inform when and how often initial research impact assessments are re-visited to ensure that they remain adequately responsive to factors that could present new potential harms or significantly influence impacts that have been previously identified:

1. ***Research workflow and production factors***: Choices made at any point along the research workflow may affect the veracity of prior impact assessments—leading to a need for re-assessment, reconsideration, and amendment. For instance, research design choices could be made that were not anticipated in the initial impact assessment (such choices might include adjusting the variables that are included in the model, choosing more complex algorithms, or grouping variables in ways that may impact specific groups). These changes may influence how a computational model performs, how it is explained, or how it impacts affected individuals and groups. Research processes are also iterative and frequently bi-directional, and this often results in the need for revision and update. For these reasons, research impact assessments must remain agile, attentive to change, and at-the-ready to evaluatively move back and forth across the decision-making pipeline as downstream actions affect upstream choices and evaluations.

2. **Environmental factors:** Changes in project-relevant social, regulatory, policy or legal environments (occurring during the time in which the research is taking place) may have a bearing on how well the resulting computational model works and on how the research outputs impact affected individuals and groups. Likewise, domain-level reforms, policy changes, or changes in data recording methods may take place in the population of concern in ways that affect whether the data used to train the model accurately portrays phenomena, populations, or related factors in an accurate manner. In the same vein, cultural or behavioral shifts may occur within affected populations that alter the underlying data distribution and hamper the predictive and explanatory efficacy of a model, which has been trained on data collected prior to such shifts. All of these alterations of environmental conditions can have a significant effect on how research practices, outputs, and results impact affected individual and communities.

## Reflect on purposes, positionality, and power

The foregoing elements of research impact assessment presuppose that the CSS researchers who undertake them also engage in reflexive practices that scrutinise the way potential perspectival limitations and power imbalances can exercise influence on the equity and integrity of research projects and on the motivations, interests, and aims that steer them. The imperative of reflecting on purposes, positionality, power makes explicit the importance of this dimension of inward-facing reflection.

All individual human beings come from unique places, experiences, and life contexts that shape their perspectives, motivations, and purposes. Reflecting on these contextual attributes is important insofar as it can help researchers understand how their viewpoints might differ from those around them and, more importantly, from those who have diverging cultural and socioeconomic backgrounds and life experiences. Identifying and probing these differences enables individual researchers to better understand how their own backgrounds, for better or worse, frame the way they see others, the way they approach and solve problems, and the way they carry out research and engage in innovation. By undertaking such efforts to recognise social position and differential privilege, they may gain a greater awareness of their own personal biases and unconscious assumptions. This then can enable them to better discern the origins of these biases and assumptions and to confront and challenge them in turn.

Social scientists have long referred to this site of self-locating reflection as "positionality" (Bourke, 2014; Kezar, 2002; Merriam et al., 2001). When researchers take their own positionalities into account, and make this explicit, they can better grasp how the influence of their respective social and cultural positions potentially creates research strengths and limitations. On the one hand, one's positionality—with respect to characteristics like   ethnicity, race, age, gender, socioeconomic status, education and training levels, values, geographical background, etc.—can have a positive effect on an individual's contributions to a research project; the uniqueness of each person's lived experience and standpoint can play a constructive role in introducing insights and understandings that other team members do not have. On the other hand, one's positionality can assume a harmful role when hidden biases and prejudices that derive from a person's background, and from differential privileges and power imbalances, creep into decision-making processes undetected and subconsciously sway the purposes, trajectories, and approaches of research projects.

When taking positionality into account, researchers should reflect on their own ***positionality matrix***. They should ask: To what extent do my personal characteristics, group identifications, socioeconomic status, educational, training, & work background, team composition, & institutional frame represent sources of power and advantage or sources of marginalisation and disadvantage? How does this positionality influence my (and my research team's) ability to identify & understand affected stakeholders and the potential impacts of my project? Answering these questions involves probing several other areas of self-ascription related to each researcher's contextual attributes (visualised in Fig. 1):

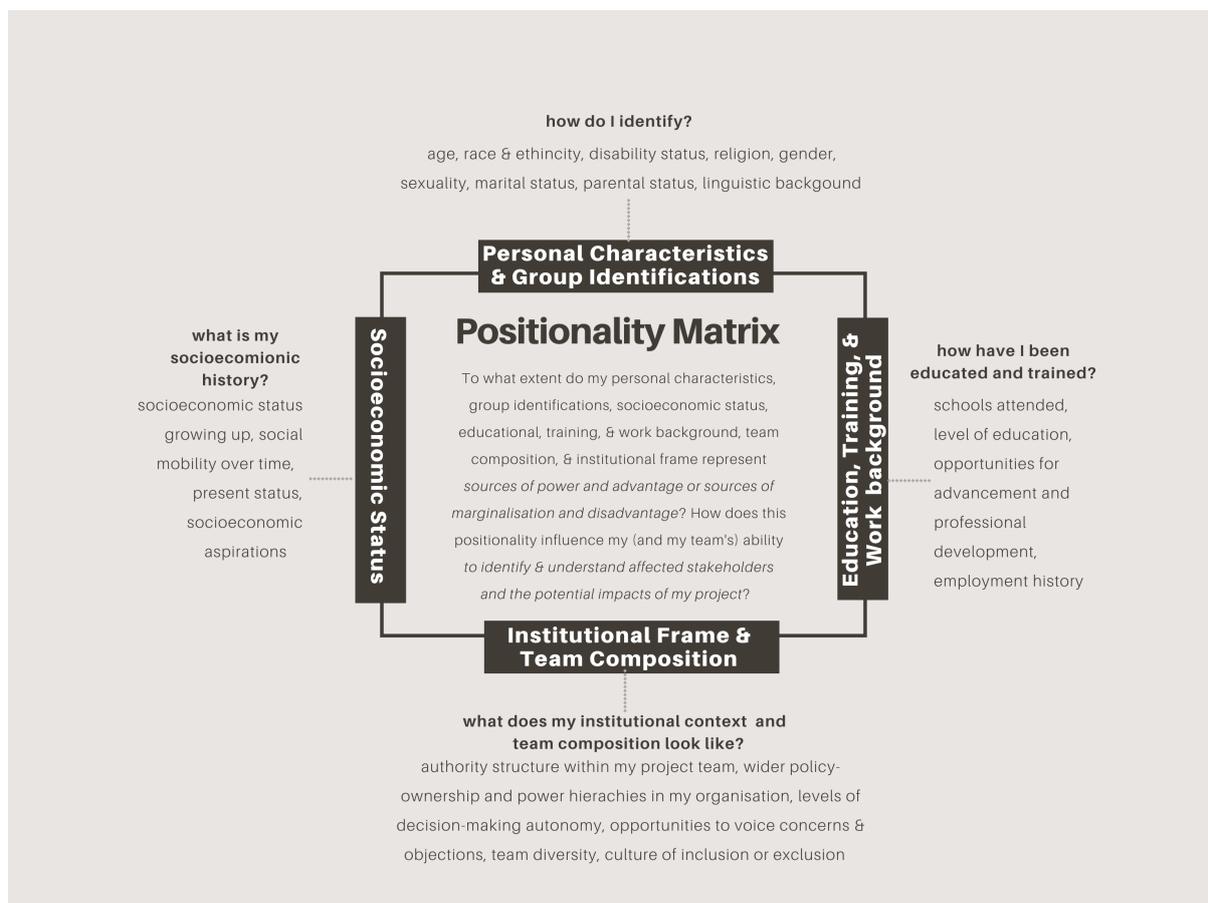

*Figure 1: Positionality Matrix (Leslie et al., 2022c)*

A solid grasp on positionality allows researchers to better interrogate and reflect on the power dynamics that could unduly influence research purposes and trajectories. Such reflections on power should involve an investigation of how power operates, and where it manifests, both across the research lifecycle and in the real-world environments in which research practices, results, and outputs are situated. Leslie et al. (2022b, pp. 52-53) propose a series of guiding questions that can be used as a reflective tool to help make potentially noxious power dynamics explicit:

> What, if any, power imbalances exist between me (or my research team) and the communities impacted by our research?
> - Do the research agendas I currently pursue reinforce or challenge these imbalances?
> - How, if at all, do these imbalances result in unjust exercises of power? Are my current activities entrenching or combating such exercises of power?

What are my interests (or my research team's interest) in collecting or procuring data and in using these to build models and answer research questions?

- How, if at all, are these interests similar to or different from the interests of those in the communities that research impacts?
- How, if at all, do any power imbalances that exist between me (or my research team) and impacted communities influence the pursuit of these interests in my (or my team's) research agendas?
- How, if at all, do I (or my research team) exploit power imbalances to pursue these interests?

What other actors hold power and influence over the research agendas I pursue and the ways I collect or procure data and build and implement models and data applications?

- How reliant am I on the data, tools, models, and digital infrastructure (connectivity, computing resources, and data assets) provided by other actors?
- What are the interests of these actors? How are they similar to or different from my interests and from those of the members of the communities impacted by my data work?
- What, if any, power imbalances exist between these actors and me (and my firm or organisation)?
- What is the history of these power imbalances? Are current policies and available resources reinforcing or contesting these imbalances?
- How, if at all, do these imbalances result in unjust exercises of power? Are current policies and available resources enabling or combating such exercises of power?

What does the institutional context of my research team look like (taking into account the authority structure within my team(s), wider policy-ownership and power hierarchies in my organisation, levels of decision-making autonomy, and opportunities to voice concerns)?

Does this institutional context enable my research practices to safeguard the public interest and to ensure that standards and governance regimes in the research ecosystem are working towards just and societally beneficial outcomes?

## Engage inclusively

While practices of inward-facing reflection on purposes, positionality, and power can strengthen the reflexivity, objectivity, and reasonableness of CSS research activities (D'Ignazio and Klein, 2020; Haraway, 1988; Harding, 1992, 1995, 2008, 2015), practices of outward-facing stakeholder engagement and community involvement can bolster a research project's legitimacy, social license, and democratic governance as well as ensure that its outputs will possess an appropriate degree of public accountability and transparency. A diligent stakeholder engagement process can help research teams to identify stakeholder salience, undertake team positionality reflection, and facilitate proportionate community involvement and input throughout the research project workflow. This process can also safeguard the equity and the contextual accuracy of impact assessments and facilitate appropriate end-to-end processes of transparent project governance by supporting their iterative revisitation and re-evaluation. Moreover, community-involving engagement processes can empower the public and the CSS community alike by introducing the transformative agency of "citizen science" into research processes (Albert et al., 2021; Sagarra et al., 2016; Tauginienė et al., 2020).

It is important to note, however, that all stakeholder engagement processes can run the risk either of being cosmetic or tokenistic tools employed to legitimate research projects without substantial and meaningful participation or of being insufficiently participatory, i.e., of being one-way information flows or nudging exercises that serve as public relations instruments (Arnstein, 1969; Tritter and McCallum, 2006). To avoid such hazards of superficiality, CSS researchers should shore up a proportionate approach to stakeholder engagement through deliberate and precise goal setting. The objectives of engagement that your research team chooses will depend on factors that divide into four categories:

1. **Assessment of risks of adverse impacts:** Stakeholder involvement in CSS research projects should be proportionate to the scope of their potential risks and hazards.

2. **Assessment of positionality**: Stakeholder involvement should address positionality limitations. For instance, in cases where the identity characteristics of research team members do not sufficiently reflect or represent significantly impacted groups, stakeholder participation can "fill gaps" in knowledge, domain expertise, and lived experience.

3. **Assessment of research needs:** Stakeholder involvement should help CSS researchers strengthen their ability to frame research questions and to tackle research problems. Researchers should explore the optimal means for community members to actively contribute to their scientific practices.

4. **Establishment of stakeholder engagement goals:** Researchers should determine engagement objectives that enable the appropriate degree of stakeholder engagement and co-production in project design, evaluation, and oversight processes. This can be done by careful selection of participation goals from a spectrum of engagement options (informing, partnering, consulting, empowering) that equip a project with a level of engagement which meets team-based assessments of risk, research need, and positionality.

Researchers may face practical challenges when drawing on these four factors to operationalise a stakeholder engagement process. For example, limits on available resources and tight timelines could be at cross-purposes with the degree of stakeholder involvement that is recommended by team-based assessments of research needs, potential hazards, and positionality limitations. Likewise, the chosen degree of appropriate public participation may be unrealistic or out-of-reach given the engagement barriers that arise from constraints on the capacity of vulnerable stakeholder groups to participate, difficulties in reaching marginalised, isolated, or socially excluded groups, and challenges to participation that are presented by digital divides or information and communication gaps between research organisations and impacted communities. In these instances, research teams should take a deliberate and reflective approach to deciding on how to balance engagement goals with practical considerations and should, at all events, make explicit the rationale behind their choices and document this.

Regardless of any potential trade-offs, researchers should prioritise the establishment of clear and explicit stakeholder engagement goals. Relevant questions to pose in establishing these goals include: *Why are we engaging with stakeholders? What do we envision the ideal purpose and the expected outcomes of engagement activities to be? How can we best draw on the insights and lived experience of participants to inform and shape our research?* To answer these questions with sufficient specificity, researchers should refer the spectrum of engagement options, keeping in mind the both engagement goals and practical constraints (Fig. 2):

## SPECTRUM OF ENGAGEMENT OPTIONS

| | DEGREE OF PARTICIPATION | MEANS OF PARTICIPATION | LEVEL OF AGENCY |
|---|---|---|---|
| **EMPOWER** | *Stakeholders are engaged with as decision-makers and are expected to gather pertinent information and be proactive in co-production.* | *Co-production exercises occur through citizens' juries, citizens' assemblies, and participatory co-design. Teams provide support for stakeholders' decision making.* | *Stakeholders exercise a high level of agency and control over agenda-setting and decision-making.* |
| **PARTNER** | *Stakeholders and teams share agency over the determination of areas of focus and decision making.* | *External input is sought out for collaboration and co-production. Stakeholders are collaborators in projects. They are engaged through focus groups.* | *Stakeholders exercise a moderate level of agency in helping to set agendas through collaborative decision-making* |
| **CONSULT** | *Stakeholders can voice their views on pre-determined areas of focus, which are considered in desicion-making.* | *Engagement occurs through online surveys or short phone interviews, door-to-door or in public spaces. Broader listening events can support consultations.* | *Stakeholder are included as sources of information input under narrow, highly controlled conditions of participation.* |
| **INFORM** | *Stakeholders are made aware of decisions and developments.* | *External input is not sought out. Information flows in one direction. This can be done through newsletters, the post, app notifications or community forums.* | *Stakeholders are treated as information subjects rather than active agents.* |

*Figure 2: Spectrum of engagement options (Leslie et al., 2022c)*

## Act transparently and responsibly

The imperative of acting transparently and responsibly enjoins CSS researchers to marshal the habits of responsible research and innovation cultivated in the CARE processes to produce research that prioritises data stewardship and that is robust, accountable, fair, non-discriminatory, explainable, reproducible, and replicable. While the mechanisms and procedures which are put in place to ensure that these normative goals are achieved will differ from project to project (based on the specific research contexts, research design, and research methods), all CSS researchers should incorporate the following priorities into their governance, self-assessment, and reporting practices:

*Full documentation of data provenance, lineage, linkage, and sourcing:* This involves keeping track of and documenting both responsible data management practices across the entire research lifecycle, from data extraction or procurement and data analysis, cleaning, and pre-processing to data use, retention, deletion, and updating (Bender & Friedman, 2018; Gebru et al., 2021; Holland et al., 2018). It also involves demonstrating that the data is ethically sourced, responsibly linked, and legally available for research purposes (Weinhardt, 2020) and making explicit measures taken to ensure data quality (source integrity and measurement accuracy, timeliness and recency, relevance, sufficiency of quantity, dataset representativeness), data integrity (attributability, consistency, completeness, contemporaneousness, traceability, and auditability) and FAIR data (Findable, Accessible, Interoperable, and Reusable).

*Full documentation of privacy, confidentiality, consent, and data protection due diligence:* This involves demonstrating that data has been handled securely and responsibly from beginning to end of the research lifecycle so that any potential breaches of confidentiality, privacy, and anonymity have been prevented and any risks of re-identification through triangulation and data linkage mitigated. Regardless of the jurisdictions of data collection and use, researchers should aim to optimally protect the rights and interests of research subjects by adhering to the highest standards of privacy preservation, data protection, and responsible data handling and storage such as those contained in the IRE 3.0 and the NESH guidelines (franzke, 2020; NESH, 2019). They should also demonstrate that they have sufficiently taken into account contextual factors in meeting the privacy expectations of observed research subjects (like who is involved in observed interactions, how and what type of information is exchanged, how sensitive it is perceived to be, and where and when such exchanges occur). Documentation should additionally include evidence that researchers have instituted proportionate protocols for attaining informed and meaningful consent that are appropriate to the specific contexts of the data extraction and use and that cohere with the reasonable expectations of targeted research subjects.

*Transparent and accountable reporting of research processes and results and appropriate publicity of datasets:* Research practices and methodological conduct should be carried out deliberately, transparently, and in accordance with recording protocols that enable the interpretability, reproducibility, and replicability of results. For prediction models, the documentation protocols presented in Transparent Reporting of a Multivariable Prediction Model for Individual Prognosis or Diagnosis (TRIPOD) provide a good starting point for best conduct guidelines in research reporting (Collins et al., 2015; Moons et al., 2015).[13] Following TRIPOD, transparent and accountable reporting should demonstrate diligent methodological conduct across all stages and elements of research. For prediction models, this includes clear descriptions of research participants, predictors, outcome variables, sample size, missing data, statistical analysis methods, model specification, model performance, model validation, model updating, and study limitations. While transparent research conduct can facilitate reproducibility and replicability, concerns about the privacy and anonymity of research subjects should also factor into how training data, models, and results are made available to the scientific community. This notwithstanding, CSS researchers should prioritise the publication of well-archived, high quality, and accessible datasets that enable the replication of results and the advancement of further research (Hollingshead et al., 2021). They should also pursue research design, analysis, and reporting in an interpretability-aware manner that prioritises process transparency, the understandability of models, and the accessibility and explainability of the rationale behind their results.

*An end-to-end process for bias self-assessment:* This should cover all research stages as well as all sources of biases that could arise in the data, in the data collection, in the data pre-processing, in the organising, categorising, describing, annotating, structuring of data (text-as-data, in particular), and in research design and execution choices. Bias self-assessment processes should cover *social, statistical, and cognitive biases* (Leslie et al., 2022e). Social bias has to do with the way that pre-existing or historical patterns of discrimination and social injustice – and the prejudices and discriminatory attitudes that correspond to such patterns – can be drawn into the research lifecycle. In particular, it relates to how these patterns and attitudes can be perpetuated, reinforced, or exacerbated through the design and development of computational models. Statistical bias refers to a systematic deviation from an expected statistical result that arises due to the influence of some additional factor. This understanding is common in observational studies where bias can arise in the process of sampling or measurement. Statistical biases can involve errors (deviations from a true state) or differences between measured or calculated values and true values. Cognitive bias refers to a systematic deviation from a norm of rationality that can occur in processes of thinking or judgement and that can lead to mental errors, misinterpretations of information, or flawed patterns of response to decision problems. An end-to-end process for bias self-assessment should move across the research lifecycle, pinpointing specific forms of social, statistical, and cognitive bias that could arise at each stage (for instance, social biases like representation bias and label bias as well as statistical biases like missing data bias and measurement bias could arise in the data pre-processing stage of a research project).

# Conclusion

This chapter has explored the spectrum of ethical challenges that CSS faces across the myriad possibilities of its application. It has further elaborated on how these challenges can be met head-on only through the adoption of habits of RRI that are instantiated in end-to-end governance mechanisms which set up practical guardrails throughout the research lifecycle. As a quintessential *social impact science*, CSS holds great promise to advance social justice, human flourishing, and biospheric sustainability. However, CSS is also an *all-too-human science*—conceived in particular social, cultural, and historical contexts and pursued amidst intractable power imbalances, structural inequities, and potential conflicts of interest. Its proponents, in both research and policymaking communities, must thus remain continuously self-critical about the role that values, interests, and power dynamics play in shaping mission-driven research. Likewise, they must vigilantly take heed of the complicated social and historical conditions surrounding the generation and construction of data as well as the way that the activities and theories of CSS researchers can function to reformat, reorganise, and shape the phenomena that they purport only to measure and analyse. Such a continuous labour of exposing and redressing the often-concealed interdependencies that exist between CSS and the social environments in which its research activities, subject matters, and outputs are embedded will only *strengthen its objectivity* and ensure that its impacts are equitable, ethical, and responsible. Such a human-centred approach will make CSS a "science with and for society" second-to-none.

---

[13] Though the TRIPOD method is intended to be applied in the medical domain, its reporting protocols are largely applicable to CSS studies.

# Annex 1: 12 Principles and Priorities of Responsible Data Innovation

The information contained below serves as background material to provide you with a means of accessing and understanding some of the existing human rights, fundamental freedoms, and value priorities that could be impacted by CSS research. A thorough review of this table and engagement of the links to the relevant Charters, Conventions, Declarations, and elaborations it contains are first steps that will help you identify the salient rights, freedoms, and values that could be affected by your research project.

| Principles and Priorities | Corresponding Rights and Freedoms with Selected Elaborations | Resources for Principles and Priorities and Corresponding Rights and Freedoms |
|---|---|---|
| **Respect for and protection of human dignity** | *All individuals are inherently and inviolably worthy of respect by mere virtue of their status as human beings. Humans should be treated as moral subjects, and not as objects to be algorithmically scored or manipulated.*<br>~<br>**-The right to human dignity, the right to life and the right to physical, mental and moral integrity**<br><br>**-The right to be informed of the fact that one is interacting with a algorithmic system rather than with a human being**<br><br>**-The right to refuse interaction with an algorithmic system whenever this could adversely impact human dignity** | **Universal Declaration of Human Rights:**<br>-Preamble, <u>Universal Declaration of Human Rights</u> – *Dignity*<br><br>**International Covenant on Civil and Political Rights:**<br>-Article 6, <u>International Covenant on Civil and Political Rights</u> – *Right to life*<br><br>**European Convention on Human Rights (ECHR):**<br>-Article 2, <u>European Convention on Human Rights</u> – *Right to life*<br><br>-Article 2, <u>'Guide on Article 2 of the European Convention on Human Rights'</u>, Council of Europe – *Right to life*<br><br>**African Commission on Human and Peoples' Rights** |
| | | 473 Resolution on the need to undertake a Study on human and peoples' rights and artificial intelligence (AI), robotics and other new and emerging technologies in Africa - ACHPR/Res. 473 |
| Interconnectivity, solidarity, and intergenerational reciprocity | *All humans are interconnected to a greater whole, which transcends time and thrives when all its constituent parts are enabled to thrive. This unbounded bond of solidarity extends from the closest relationship between kin to the living totality of the biospheric whole. Membership in this greater community also places a responsibility on the present generation to take account of the well-being and flourishing of future generations. Intergenerational reciprocity involves looking backward in considering the wisdom and learning of past generations and looking forward in considering the rights and well-being of lives not yet lived (two, four, seven, or more generations in the future).*<br>~<br><br>-The right of future generations to due moral regard and consideration<br><br>- **Kaitiakitanga** (Maori): The responsibility to ensure sustainable futures for the biosphere and for people, families, communities, and humanity<br><br>- **Manaakitanga** (Maori): The responsibility to extend care, compassion, hospitality, and generosity to all others including strangers and the environment. Shared *Manaakitanga* supports well-being, dignity, and the stewardship of healthful and spiritual living.<br><br>-The Seventh Generation Principle (Haudenosaunee Confederacy, Iroquois): Give regard to the well-being of the seventh generation ahead of you in your practices, works, actions, and deliberations and draw on the experience and wisdom of the seventh generation that came before<br><br>-The values of *Ubuntu* (Sub-Saharan Africa): Ethical life is measured by the meaningful relationships formed by each individual with an interconnected and interdependent whole of people, community, and environment. One's humanity is affirmed by connecting with and taking care of others and by recognising their dignity in works, deliberations, and deeds. | **UNESCO:**<br>-III.1 Values, Recommendation on the Ethics of Artificial Intelligence, *Living in peaceful, just and interconnected societies*<br><br>**Other resources:**<br><br>The Maori Report, **Independent Maori Statutory Body**<br><br>Treaty of Waitangi/Te Tiriti and Māori Ethics Guidelines for: AI, Algorithms, Data and IOT, **2020**<br><br>The World People's Conference on Climate Change and the Rights of Mother Earth, **Bolivia 2010**<br><br>The Constitution of the Iroquois Nations, **1916**<br><br>What is Ubuntu?, **Desmond Tutu 2013**<br><br>I am because you are, **Michael Onyebuchi Eze, UNESCO 2011** |
| Environmental flourishing, sustainability, and the rights of the biosphere | *All humans draw oxygen from the Earth's air, draw nourishment from its soil, and live as interconnected parts of a living biospheric community. The interrelated organisms of this unbounded community share a common origin, a common history, and a common ecological fate. Members of humanity, as benefactors and inheritors of such a circle of life and of the life-giving gifts of the earth, should seek practices of living that secure environmental flourishing,* | **UNESCO:**<br>-III.1 Values, Recommendation on the Ethics of Artificial Intelligence, *Environment and ecosystem flourishing*<br><br>**Other resources:**<br><br>The Constitution of Ecuador, **2008** |
| | *sustainability, and the rights of the biosphere. These practices of living should aim for a harmony and balance with the interdependent ecologies of the biosphere in solidarity with it. They should also respect nature's right to flourish, to endure, and to regenerate life without harmful anthropogenic influence. All people involved in computational research and data innovation lifecycles should prioritise environmental flourishing, sustainability, and the rights of the biosphere, ensuring that they use the affordances of technology to do battle with climate change and biodiversity drain rather than contribute to them.*<br>~<br><br>-The right of *Pachamama*: 'Nature or *Pachamama*, where life is reproduced and exists, has the right to exist, persist, maintain and regenerate its vital cycles, structure, functions and its processes of evolution'. (Article 1, Constitution of Ecuador)<br><br>-*Sumak kawsay* (Quechua), *suma qamaña* (Aymara), *buen vivir* (Spanish): "living well" or "collective well-being" but also the priority of a shared pursuit of the fullness, creativity, harmony, and flourishing of human and biospheric life.<br><br>- *Kaitiakitanga* (Maori): The responsibility to ensure sustainable futures for the biosphere and for people, families, communities, and humanity<br><br>- 'Environmental Justice affirms the sacredness of Mother Earth, ecological unity and the interdependence of all species, and the right to be free from ecological destruction'. (First National People of Colour Environmental Leadership Summit) | 17 Principles of Environmental Justice, **First National People of Colour Environmental Leadership Summit 1991**<br><br>Bali Principles of Climate Justice**, 2002**<br><br>The Maori Report**, Independent Maori Statutory Body**<br><br>Treaty of Waitangi/Te Tiriti and Māori Ethics Guidelines for: AI, Algorithms, Data and IOT, **2020**<br><br>The World People's Conference on Climate Change and the Rights of Mother Earth, **Bolivia 2010**<br><br>The Albuquerque Declaration, **Native People-Native Homelands Climate Change Workshop-Summit, Albuquerque, New Mexico, 1998** |
| **Protection of human freedom and autonomy** | *Humans should be empowered to determine in an informed and autonomous manner if, when, and how algorithmic and data-intensive systems are to be used. These systems should not be employed to condition or control humans, but should rather enrich their capabilities.*<br><br>~<br><br>-The right to liberty and security<br><br>-The right to human autonomy and self-determination<br><br>-The right not to be subject to a decision based solely on automated processing when this produces legal effects on groups or similarly significantly affects individuals<br><br>-The right to effectively contest and challenge decisions informed and/or made by an AI | **Universal Declaration of Human Rights:**<br> -Article 3, Universal Declaration of Human Rights – *Right to life, liberty, and the security of person*<br><br>-Article 18, Universal Declaration of Human Rights – *Right to freedom of thought, conscience, and religion*<br><br>-Article 19, Universal Declaration of Human Rights – *Right to freedom of opinion and expression*<br><br>**African Commission on Human and Peoples' Rights**<br>473 Resolution on the need to undertake a Study on human and peoples' rights and artificial intelligence (AI), robotics and other new and emerging technologies in Africa - ACHPR/Res. 473 |
| | system and to demand that such decisions be reviewed by a person<br><br>-The right to freely decide to be excluded from AI-enabled manipulation, individualised profiling, and predictions. This also applies to cases of non-personal data processing<br><br>-The right to have the opportunity, when it is not overridden by competing legitimate grounds, to choose to have contact with a human being rather than a robot | **International Covenant on Civil and Political Rights:**<br> -Article 9, <u>International Covenant on Civil and Political Rights</u> – *Right to liberty and security of person*<br><br>-Article 18, <u>International Covenant on Civil and Political Rights</u> – *Right to freedom of thought, conscience, and religion*<br><br>-Article 19, <u>International Covenant on Civil and Political Rights</u> – *Freedom of expression*<br><br>**European Convention on Human Rights (ECHR):**<br> -Article 5, <u>European Convention on Human Rights</u> – Right to liberty and security<br><br>-Article 5, <u>'Guide on Article 5 of the European Convention on Human Rights'</u>, Council of Europe – *Right to liberty and security*<br><br>-Article 9, <u>European Convention on Human Rights</u> – *Freedom of thought, conscience, and religion*<br><br>-Article 9, <u>'Guide on Article 9 of the European Convention on Human Rights'</u>, Council of Europe – *Freedom of thought, conscience, and religion*<br><br>-Article 10, <u>European Convention on Human Rights</u> – *Freedom of expression*<br><br>*-Article 10,* <u>*'Guide on Article 10 of the European Convention on Human Rights'*</u>, Council of Europe – *Freedom of expression* |
| **Prevention of harm and protection of the right to life and physical, psychological, and moral integrity** | *The physical and mental integrity of humans and the sustainability of the biosphere must be protected, and additional safeguards must be put in place to protect the vulnerable. Algorithmic and data-intensive systems must not be permitted to adversely impact human well-being or planetary health.*<br><br>~<br><br>-The right to life and the right to physical and mental integrity<br><br>-The right to the protection of the environment<br><br>-The right to sustainability of the community and biosphere | **European Convention on Human Rights (ECHR):**<br> -Article 2, <u>European Convention on Human Rights</u> – *Right to life*<br><br>-Article 2, <u>'Guide on Article 2 of the European Convention on Human Rights'</u>, Council of Europe – *Right to life* |
| **Non-discrimination, fairness, and equality** | *All humans possess the right to non-discrimination and the right to equality and equal treatment under the law. Algorithmic and data-intensive systems must be designed to be fair, equitable, and inclusive in their beneficial impacts and in the distribution of their risks.*<br><br>~ | **Universal Declaration of Human Rights:**<br> -Article 7, <u>Universal Declaration of Human Rights</u> – *Equality before the law*<br><br>**African Commission on Human and Peoples' Rights** |
| | -The right to non-discrimination, including intersectional discrimination<br><br>-The right to non-discrimination and the right to equal treatment. This right must be ensured in relation to the entire lifecycle of an algorithmic system (design, development, implementation, and use), as well as to the human choices concerning its design, adoption, and use, whether used in the public or private sector. | 473 Resolution on the need to undertake a Study on human and peoples' rights and artificial intelligence (AI), robotics and other new and emerging technologies in Africa - ACHPR/Res. 473<br><br>**International Covenant on Civil and Political Rights:**<br>-Article 6, International Covenant on Civil and Political Rights – *Right to life*<br><br>-Article 26, International Covenant on Civil and Political Rights – *Non-discrimination*<br><br>**European Convention on Human Rights (ECHR):**<br>-Protocol No. 12, European Convention on Human Rights<br><br>-Article 14, European Convention on Human Rights – *Prohibition of discrimination*<br><br>-Article 14 and Article 12 of Protocol No. 12, 'Guide on Article 14 of the European Convention on Human Right and on Article 1 of Protocol No. 12 to the Convention', Council of Europe – *Prohibition of discrimination*<br><br>**Office of the United Nations High Commissioner for Human Rights:**<br>-OHCHR, International Convention on the Elimination of All Forms of Racial Discrimination<br><br>-OHCHR, Convention on the Elimination of All Forms of Discrimination against Women |
| **Rights of Indigenous Peoples and Indigenous Data Sovereignty** | *Indigenous peoples have a right to self-determination, to recognition of equal standing, and to remedy and reparation for the historical and systemic denial of their rights. These rights should be contextualised in accordance with the unique sociocultural histories and lived experience of the Indigenous people to whom such rights apply. Indigenous peoples also have a right to control data from and about their communities, activities, and lands and to shape the way these are collected and used. This encompasses both collective rights of benefit, access, ownership, and control and individual data-related rights and freedoms like rights to privacy and dignity.*<br><br>~<br><br>-The rights to the restoration of equality, reparation, and self-determination<br><br>- *Rangatiratanga* (Maori): The empowering unity of a self-determining and sovereign community that is bound together by the reciprocal involvement of leadership and community members in collective governance, | **The United Nations**<br>- United Nations Declaration on the Rights of Indigenous Peoples<br><br>The Maori Report, **Independent Maori Statutory Body, 2016**<br><br>Treaty of Waitangi/Te Tiriti and Māori Ethics Guidelines for: AI, Algorithms, Data and IOT, **2020**<br><br>Compendium of Māori Data Sovereignty, **2022**<br><br>Barunga Statement, **Aboriginal and Torres Strait Islander Peoples 1988**<br><br>Uluru Statement from the Heart, **Aboriginal and Torres Strait Islander Peoples, National Constitutional Convention 2017**<br><br>Idle No More Movement, **First Nations of Canada 2012** |
| | problem solving, and the articulation of shared goals and visions<br><br>- *Makarrata* (Aboriginal and Torres Strait Islander): The coming together after a struggle, confronting harms done, truth telling, righting the wrongs of the past, and restoring peace, solidarity, and community | The CARE Principles for Indigenous Data Governance, **2020** |
| **Data protection and the right to respect of private and family life** | *The design and use of algorithmic and data-intensive systems that rely on the processing of personal data must secure a person's right to respect for private and family life, including the individual's right to control their own data. Informed, freely given, and unambiguous consent must play a role in this.*<br>~<br><br>-**The right to respect for private and family life and the protection of personal data**<br><br>-**The right to physical, psychological, and moral integrity in light of algorithm-based profiling and emotion/personality recognition**<br><br>-**All the rights enshrined in Convention 108+ of the Council of Europe and in its modernised version, and in particular with regard to AI-based profiling and location tracking** | **Universal Declaration of Human Rights:**<br> -Article 12, Universal Declaration of Human Rights – *Right to respect for privacy, family, home, or correspondence*<br><br>**African Commission on Human and Peoples' Rights**<br>473 Resolution on the need to undertake a Study on human and peoples' rights and artificial intelligence (AI), robotics and other new and emerging technologies in Africa - ACHPR/Res. 473<br><br>**African Union**<br>-African Union Convention on Cyber Security and Personal Data Protection, 2014<br><br>**European Convention on Human Rights (ECHR):**<br> -Article 8, European Convention on Human Rights – *Right to respect for private and family life*<br><br>-Article 8, 'Guide on Article 8 of the European Convention on Human Rights. Right to respect for private and family life, home and correspondence', Council of Europe – *Right to respect for private and family life* |
| **Economic and social rights** | *Individuals must have access to the material means needed to participate fully in work life, social life, and creative life, and in the conduct of public affairs, through the provision of proper education, adequate living and working standards, health, safety, and social security. This means that algorithmic and data-intensive systems should not infringe upon individuals' rights to work, to just, safe, and healthy working conditions, to social security, to the protection of health, and to social and medical assistance.*<br>~<br><br>-**The right to just working conditions, the right to safe and healthy working conditions, the right to organise, the right to social security, and the rights to the protection of health and to social and medical assistance** | **African Union**<br>Digital Transformation Strategy for Africa (2020-2030)<br><br>**Universal Declaration of Human Rights:**<br> -Article 3, Universal Declaration of Human Rights – *Right to life, liberty, and the security of person*<br><br>-Article 12, Universal Declaration of Human Rights – *Right to private home life*<br><br>-Article 22, Universal Declaration of Human Rights – *Right to social security*<br><br>-Article 22, Universal Declaration of Human Rights – *Workers' rights*<br><br>**International Covenant on Economic, Social and Cultural Rights:**<br> -Article 6, International Covenant on Economic, Social, and Cultural Rights – *The right to work* |
| | | -Article 7, <u>International Covenant on Economic, Social, and Cultural Rights</u> – *Right to just and favourable conditions of work*<br><br>-Article 8, <u>International Covenant on Economic, Social, and Cultural Rights</u> – *Right to organise*<br><br>-Article 9, <u>International Covenant on Economic, Social, and Cultural Rights</u> – *Right to social security* |
| **Accountability and effective remedy** | *Accountability demands that the onus of justifying outcomes that have been influenced by data-driven and algorithmic systems be placed on the shoulders of the human creators and users of those systems. This means that it is essential to establish a continuous chain of human responsibility across the whole data innovation lifecycle. Making sure that accountability is effective from end to end necessitates that no gaps be permitted in the answerability of responsible human authorities from first steps of the design of a system to its deprovisioning. Accountability also entails that every step of the process of designing and implementing the system is accessible for audit, oversight, and review. Where a system harms people, they have a right to actionable recourse and effective remedy, so that responsible parties can be held accountable.*<br><br>~<br><br>**-The right to an effective remedy for violation of rights and freedoms. This should also include the right to effective and accessible remedies whenever the development or use of algorithmic and data-intensive systems by private or public entities causes unjust harm or breaches an individual's legally protected rights.** | **Universal Declaration of Human Rights:**<br> -Article 8, <u>Universal Declaration of Human Rights</u> – *Right to an effective remedy*<br><br>**International Covenant on Civil and Political Rights:**<br> -Article 2, <u>International Covenant on Civil and Political Rights</u> – *Right to effective remedy*<br><br>**European Convention on Human Rights (ECHR):**<br> -Article 13, <u>European Convention on Human Rights</u> – *Right to an effective remedy*<br><br>-Article 13, <u>'Guide on Article 13 of the European Convention on Human Rights.'</u>, Council of Europe – *Right to an effective remedy* |
| **Democracy** | *Individuals should enjoy the ability to freely form bonds of social cohesion, human connection, and solidarity through inclusive and regular democratic participation, whether in political life, work life, or social life. This requires informational plurality, the free and equitable flow of the legitimate and valid forms of information, and the protection of freedoms of expression, assembly, and association.*<br><br>~<br><br>**-The right to freedoms of expression, assembly, and association**<br><br>**-The right to vote and to be elected, the right to free and fair elections, and in particular universal, equal and free suffrage, including equality of opportunities and the freedom of voters to form an opinion. In this regard, individuals should not be subjected to any deception or manipulation.** | **Universal Declaration of Human Rights:**<br> -Article 19, <u>Universal Declaration of Human Rights</u> – *Right to freedom of opinion and expression*<br><br>-Article 20, <u>Universal Declaration of Human Rights</u> – *Right to freedom of peaceful assembly and association*<br><br>**International Covenant on Civil and Political Rights:**<br> -Article 19, <u>International Covenant on Civil and Political Rights</u> – *Freedom of expression*<br><br>-Article 21, <u>International Covenant on Civil and Political Rights</u> – *Freedom of assembly*<br><br>-Article 22, <u>International Covenant on Civil and Political Rights</u> – *Freedom of association*<br><br>-Article 25, <u>International Covenant on Civil and Political Rights</u> – *Right to participate in public affairs, good governance, and elections* |
| | -The right to (diverse) information, free discourse, and access to plurality of ideas and perspectives<br><br>-The right to good governance | **European Convention on Human Rights (ECHR):**<br> -Article 3 of Protocol No.1, <u>European Convention on Human Rights</u> – *Right to free elections*<br><br>- Article 3 of Protocol No. 1, <u>Guide on Article 3 of Protocol No. 1 to the European Convention of Human Rights</u> – *Right to free elections*<br><br>-Article 10, <u>European Convention on Human Rights</u> – *Freedom of expression*<br><br>-*Article 10,* <u>'Guide on Article 10 of the European Convention on Human Rights'</u>, Council of Europe – *Freedom of expression*<br><br>-Article 11, <u>European Convention on Human Rights</u> – *Freedom of assembly and association*<br><br>-Article 11, <u>'Guide on Article 11 of the European Convention on Human Rights'</u>, Council of Europe – *Freedom of assembly and association* |
| **Rule of law** | *Algorithmic and data-intensive systems must not undermine judicial independence, effective remedy, the right to a fair trial, due process, or impartiality. To ensure this, the transparency, integrity, and fairness of the data and data processing methods must be secured.*<br>*~*<br><br>**-The right to a fair trial and due process. This should also include the possibility of receiving insight into and challenging algorithm-informed decisions in the context of law enforcement or justice, including the right to review of such decisions by a human. The essential requirements that secure impacted individuals' access to the right of a fair trial must also be met equality of arms, right to a natural judge established by law, the right to an independent and impartial tribunal, and respect for the adversarial process.**<br><br>**-The right to judicial independence and impartiality, and the right to legal assistance**<br><br>**-The right to an effective remedy, also in cases of unlawful harm or breach an individual's human rights in the context of AI and data-intensive systems** | **Universal Declaration of Human Rights:**<br>-Article 8, <u>Universal Declaration of Human Rights</u> – *Right to an effective remedy*<br><br>- Article 10, <u>Universal Declaration of Human Rights</u> – *Right to a fair trial*<br><br>**International Covenant on Civil and Political Rights:**<br> -Article 2, <u>International Covenant on Civil and Political Rights</u> – *Right to effective remedy*<br><br>-Article 14, <u>International Covenant on Civil and Political Rights</u> – *Right to fair trial*<br><br>**European Convention on Human Rights (ECHR):**<br> -Article 6, <u>European Convention on Human Rights</u> – *Right to a fair trial*<br><br>-Article 6, <u>'Guide on Article 6 of the European Convention on Human Rights.'</u>, Council of Europe – *Right to a fair trial*<br><br>-Article 13, <u>European Convention on Human Rights</u> – *Right to an effective remedy*<br><br>-Article 13, <u>'Guide on Article 13 of the European Convention on Human Rights.'</u>, Council of Europe – *Right to an effective remedy* |